# The Competitiveness of Renewables: An Analysis of Magnitude, Geography, and Drivers

Pascal Fröhlich[1, a], Maximilian Bernecker[1, b], Felix Müsgens[1, c]

[1] *Chair of Energy Economics Brandenburg University of Technology,*

*Siemens-Halske-Ring 13, 03046 Cottbus, Germany*

[a] *pascal.froehlich@b-tu.de, (corresponding author)*

[b] maximilian.bernecker@b-tu.de

[c] felix.muesgens@b-tu.de

**Abstract**

While renewable energy sources are the fastest-growing electricity generation technology globally, their competitiveness is still the subject of controversy. This paper presents an electricity system model for investment and dispatch to determine the cost-optimal shares of renewable energy sources. We compute and analyse renewable generation shares in market equilibrium for Germany and Texas, using annual data for 2015 to 2024, and in five-year intervals for 2030 to 2050. Furthermore, we identify the key drivers of the renewable competitiveness and quantify their contribution through parameter variations. Our results show that renewable generation achieves considerable market shares even without subsidies. In Germany, the increase in renewable generation is primarily driven by $CO_2$ pricing, complemented by declining investment costs for renewable technologies. In Texas, solar PV is part of the cost-optimal system, even in the absence of $CO_2$ pricing and despite low natural gas prices.

# 1 INTRODUCTION

Globally, renewable energy sources (RES) are the fastest growing electricity generation technology, driven in particular by solar photovoltaics (PV) and wind turbines [1]. In 2025, wind and solar capacity additions exceeded 800 GW, with an expected annual generation of 1.000 TWh from these new additions alone [2]. Despite this growth, the competitiveness of renewable electricity generation is controversial, both in academia and beyond.

On the one hand, international studies indicate that RES have become highly cost-competitive. Recent reports by IRENA, the IEA, and Fraunhofer ISE show that the "Levelised Cost of Electricity" (LCOE) (the most common indicator used to compare the cost competitiveness of electricity-generating technologies) of wind and solar power have declined substantially and are increasingly below that of conventional generation [1,3,4]. Similar conclusions are drawn in [5], although the cost advantages vary by region. This is supported by technology-specific analyses, e.g. [6] and [7], who analyse auction bids for offshore wind power projects and conclude that several projects are assumed to cover all their costs on the wholesale market.

On the other hand, the literature shows that market values decrease with increasing penetration levels of renewables [8]. Furthermore, researchers maintain that, due to the intermittent nature of their generation, backup capacity from conventional power plants or storage technology is required [9], and forecast renewable generation to be uncompetitive for years to come [10,11]. Some argue that governments must take additional measures to ensure the profitable integration of renewable energy sources into the market at least up to the year 2030 [12].

This controversy is still ongoing despite the large-capacity additions for two main reasons. First, the historical roll-out of wind and solar PV was mostly driven by support payments. Hence, their impressive observed empirical roll-out cannot prove competitiveness. Investment volumes are often determined by explicit policy targets, such as the installed capacity of RES (in MW), for example China's target of 1,200 GW combined solar and wind capacity by 2030 under the 14th Five-Year Plan, and Germany's targets of 115 GW onshore wind and 215 GW solar PV by 2030 (EEG 2023); or the amount of electricity generated (in MWh), such as Spain's objective of achieving 81% renewable electricity by 2030 under its National Energy and Climate Plan (NECP). Second, comparing LCOE between RES and conventional technologies ignores that RES availability follows weather conditions. This causes differences on the revenue side, which also drive competitiveness. Hence, an analysis of RES competitiveness requires looking beyond installed capacities and cost metrics. There is a lack of studies quantifying the shares of RES capacity that would be deployed in a cost-optimal electricity system without policy support instruments. Furthermore, it is unclear whether renewable technologies, e.g. solar and wind, will require subsidies forever or when and to what extent they can be considered a cost-competitive part of future energy systems.

This paper presents a model to calculate a counterfactual expansion of RES without subsidies. We develop a capacity expansion model that computes the cost-optimal electricity system based on annualised investment and generation costs. The endogenous deployment of generation capacity captures system-level interactions, including time-varying balancing requirements, storage interactions, renewable curtailment, cannibalisation, and weather-driven capacity factor correlations. This allows us to model technology competition and assess the competitiveness of renewables in an electricity-market-consistent framework.

Our study analyses the cost-optimal shares of RES for two different systems: Germany and Texas. We chose these systems because they differ in RES potential, regulation, fuel prices, and continent. Our analysis encompasses a historical horizon with a yearly resolution from 2015 to 2024, and a future horizon in five-year intervals from 2030 to 2050. Therefore, we build a database with historical and extrapolated future investment costs, fuel prices, weather-based capacity factors, export and import quantities, and electricity demand time series. Our results quantify the historical and future share of RES capacity in a cost-optimal electricity system without policy support instruments. Furthermore, our study contributes to the literature by outlining the systemic differences over time and by identifying the key factors driving the actual and future competitiveness of RES.

We find that, even in the absence of subsidy schemes, historical electricity generation from RES, particularly solar PV and wind turbines, would have achieved substantial market shares, exceeding 50% in Germany and approaching 10% in Texas. In the case of Germany, the main driver of RES profitability is a substantial increase in the price of $CO_2$ emissions, in combination with a continued decline in investment costs for renewables. In our study, these two factors have the biggest influence on renewable deployment. In contrast, in the Texan electricity market, RES face more headwinds as the negative externality of carbon dioxide emissions is not internalised and natural gas prices are relatively low. Nonetheless, solar PV would still have been part of a cost-optimal generation mix. Looking ahead to 2050, both Germany and Texas are projected to achieve substantial increases in RES shares, around 90% and 60% respectively. However, the exponential increase in marginal $CO_2$ abatement cost hinders the German system's fully achieving market-based climate neutrality by 2050. For Texas, we find that a significant share of fossil-based generation remains competitive. Overall, renewables are expected to become a long-term, central, and economically competitive part in both electricity systems. Major factors driving the future competitiveness of RES are the expectation of declining investment costs, and rising $CO_2$ prices.

Our work demonstrates that RES (solar and wind) have become competitive and have been integral parts of cost-optimal capacity mixes in both systems since 2021. Our results are relevant for policy makers as they indicate how far renewable expansion can be driven by market forces alone; and where targeted subsidy schemes remain necessary, highlight the effectiveness of climate policy instruments such as $CO_2$ pricing in shaping low carbon cost-optimal systems; and support the design and adaption of future subsidy frameworks by identifying the point at which marginal decarbonisation becomes increasingly costly.

# 2 LITERATURE REVIEW

In this section we provide an overview of the relevant literature in four steps. First, we discuss how subsidies are defined in the literature, providing the conceptual basis for the subsequent analysis. Second, we review studies that apply energy system models to the cost-optimal investment of RES, with particular treatment of subsidies in these approaches. Third, we summarise the literature on the competitiveness of RES and the conditions under which market-based deployment is considered feasible. Finally, we synthesise the most important aspects from the previous sections and identify the resulting research gap.

## 2.1 Definition and Use of the Term 'Subsidy'

Since our work focuses on competitive, i.e. subsidy-free, market shares for RES, we start by defining the term 'subsidy'. The literature provides various, often contradictory definitions. The World Trade Organization (WTO) applies a narrow legal definition, restricting subsidies to financial contributions from governments that confer a benefit. Under this framework, only explicit financial transfers or revenue foregone are considered subsidies. Neither payments from consumers to producers, as e.g. have driven the German RES roll-out until recently[1], nor missing price signals for externalities (e.g. not pricing $CO_2$ emissions) are included [13]. In contrast to the WTO, the International Monetary Fund (IMF) adopts a broader economic definition, including the failure to price negative externalities, particularly climate damage, as implicit subsidies [14]. This approach implies that ignoring the costs of $CO_2$ emissions constitute a subsidy to fossil-based generation. The European Union (EU) takes the middle ground between the two previous definitions. EU State aid law defines subsidies as selective advantages granted from state resources that distort competition.

In contrast to these more policy-oriented discussions, we follow the classical economic textbook definition, e.g. from [15], that a subsidy is any policy-induced deviation from wholesale electricity market revenues that lowers the investment or production cost of a technology. This comprises both direct payments, e.g. a market premia or feed-in tariffs, as well as indirect payments such as tax credits or grid connection costs. It is less clear-cut whether not paying for negative externalities, or governments stabilising revenue streams, for example, should be considered subsidies. The former is especially relevant in the context of our paper – namely whether conventional power plants should pay for their carbon dioxide emissions in a subsidy-free benchmark – and what the correct cost (in €/t$CO_2$) would be. In our empirical calculations, we consider $CO_2$ prices as a cost component in Germany, where the European ETS prices $CO_2$, but do not include them in the analysis in Texas, where no such mechanism is in place.

[1] The Renewable Energy Sources Act (EEG) imposed a levy on electricity consumption to cover the additional cost of renewable electricity generation. While this system was changed in 2022, up to that point, no direct government transfers were involved and hence, no subsidies paid, following this narrow definition.

## 2.2 Energy System Modelling to Optimise RES Investment

A large number of publications use energy system models (ESMs) to analyse the cost-optimal deployment of renewable electricity generation. These models typically minimise total system cost subject to demand, technical, and policy constraints, and are widely used to assess long-term investment pathways. Early applications include optimisation frameworks such as Balmorel, which determine the optimal capacity expansion under exogenously given policy assumptions [16]. More recent models extended this approach to high-resolution temporal and spatial settings, enabling detailed representation of variable RES, storage and transmission.

In most ESM studies, subsidies are implemented exogenously as either: technology-specific investment cost reductions [17], guaranteed revenue sources (e.g. feed-in tariffs) [18], or deployment constraints derived from policy targets [19,20]. As a result, these models typically do not endogenously assess whether RES investments are profitable at market prices, but rather assume support schemes as given [21]. Only a limited number of studies analyse market-based investment conditions within ESMs. For example, [6] assess offshore wind competitiveness in mature markets without subsidies, combining energy system modelling with revenue analysis to show under which market conditions subsidy-free deployment becomes viable. Earlier ESM-based work by [22] compares renewable investment pathways with and without financial support up to 2030 and concludes, under the cost assumptions prevailing at the time, that even with $CO_2$ constraints, renewables would remain economically unattractive without dedicated subsidies. In contrast, a more recent modelling work for the US power sector from [23] analyses renewable competitiveness between 2020 and 2050 under no-subsidy and carbon tax scenarios, explicitly capturing the interaction between market value and increasing generation shares. This study highlights the significant change in the cost competitiveness of renewable energies compared to previous assessments. Against this background, our contribution differs from the existing ESM literature by explicitly quantifying the cost-optimal overall electricity system, based on endogenous investment decisions. This allows us to bridge the gap between cost-optimal system design and real-world investment feasibility based on pure market competition.

## 2.3 Competitiveness of RES

To achieve the net zero emissions target, the global energy transition requires structural changes in electricity markets, infrastructure, and policy instruments [24]. In the past, subsidies were the main driver behind the growth of solar PV and wind technologies, in both Germany and Texas. However, falling levelised costs of electricity are now allowing an increasingly market-based integration. According to the International Renewable Energy Agency (IRENA), utility-scale solar PV and onshore wind are now, on average, the lowest-cost source of new electricity generation globally. Recent projects frequently undercut the generation costs of existing fossil fuel plants [3]. [25] show that both technologies have become more competitive due to falling investment costs and rising $CO_2$ prices. [6], [7], and [26] underline that investments without subsidies are already realistic in offshore wind markets in Germany. Complementary evidence from BloombergNEF shows that a growing share of new wind and solar capacity in Europe and the United States is financed on a merchant basis, relying primarily on wholesale market revenues and corporate power purchase agreements rather than public subsidies [27]. Texas offers a practical perspective on market-driven renewable systems. The ERCOT grid is based on an energy-only design without capacity payments. As [28] shows, already by 2010, this mechanism promoted investment without tax credit in wind and solar energy. [29] highlights that systematic infrastructure investments, particularly in the transmission grid, are crucial to maintaining the RES competitiveness.

At the same time, the literature cautions that competitiveness is highly context-dependent. On the one hand, the European perspective shown by [12] indicates that RES projects are not automatically profitable under current national expansion targets. Only consistent $CO_2$ pricing will result in market-based returns. In addition, [30] argues that the successful integration of RES depends on sufficient transmission capacity to prevent localised price collapses, which directly undermines their competitiveness. On the other hand, [31] emphasises that EU-wide climate targets remain achievable despite the ambitious transformation, but only as long as markets are designed efficiently and the competitiveness of renewables is not restricted by regulatory disruptions.

## 2.4 Summary of the Research Gap

Our literature review presented differences in 'subsidy' definitions, discussing in particular the role of renewable support and climate externalities. We follow a broad definition of a 'subsidy'. Furthermore, our literature review discussed energy system models dealing with renewable investments, and the future competitiveness of wind and solar energy. Our review found that most ESM studies implement subsidies exogenously through reduced investment costs, guaranteed revenues, or policy-driven deployment targets. As a result, these models typically include subsidies when optimising system costs and thus do not answer the question as to what extent RES would already be part of a cost-efficient system without subsidies. At the same time, growing evidence from the literature on renewable competitiveness shows that declining costs have improved real-world market integration. Studies on Germany and Texas, as well as reports by IRENA and BloombergNEF, show that subsidy-free or near-subsidy-free renewable projects become increasingly feasible.

However, a study examining the historical expansion of renewables in the absence of subsidies for RES – or on a pure cost basis – is missing. Therefore, our work proposes a retrospective review in a field that has predominantly focused on future time periods.

We contribute to the existing literature by addressing the following points:

1. We investigate the historical expansion of renewables on a market-basis from 2015 to 2024.
2. We quantify the historical shares of solar PV and wind power as a part of a cost-optimal system.

3. We determine the most influential factors that affect the renewable expansion.
4. We assess cost optimal RES shares in the long-term future from 2030 to 2050.

# 3 METHODOLOGY AND PARAMETRISATION

In this section, we first introduce our linear investment and dispatch model to compute the electricity system's optimal generation mix. We present the mathematical formulations of the model, in particular the objective function (to minimise the cost of electricity production) and key constraints (energy balance constraint, etc.). The model is formulated as a linear partial equilibrium model of the respective electricity market and adopts an annualised greenfield approach for each market and year. Second, we outline the case studies for Germany and Texas with parameters and assumptions. Finally, we discuss the main limitations of the modelling approach and assumptions.

## 3.1 Mathematical Formulations

The model developed for this study is based on the EM.Power Invest model, which is presented in [32]. The GAMS code used for this analysis, along with the input data, is available in the public GitHub repository: https://github.com/froehpas/Competitiveness_of_RES. We start by introducing the model's nomenclature, i.e. the main indices, parameters, and variables. This is followed by a description of our investment and dispatch model's equations.

**Model Nomenclature**

**Indices and Sets**

| | |
|---|---|
| $hy$ | Hydro generation units |
| $n$ $(alias\ nn)$ | Nodes (Germany n = 1, Texas n = 7) |
| $p$ | Group of conventional power plants |
| $r$ | Renewable energy sources |
| $r^{disp}$ | Dispatchable renewables |
| $r^{ndisp}$ | Non-dispatchable renewables |
| $s$ | Storage technologies |
| $t$ | Time steps (hourly) |
| $y$ | Year |

**Parameters**

| | |
|---|---|
| $avail_{p,y,t}$ | Availability of conventional capacity [%] |
| $avail_{n,r,y,t}$ | Availability of renewable capacity [%] |
| $avail_{s}$ | Availability of storage capacity [%] |
| $ccurt^{Gen}_{r^{curt}}$ | Variable costs for renewable curtailment [EUR/MWh] |
| $cinv_{p,y}$ | Annualised investment costs for conventional power plants [EUR/MW] |
| $cinv_{r,y}$ | Annualised investment costs for renewable energy sources [EUR/MW] |
| $cinv_{s,y}$ | Annualised investment costs for battery capacities [EUR/MW] |
| $cramp_{p,y}$ | Ramping costs [EUR/MW] |
| $load_{n,y,t}$ | Load from demand and net export [MWh] |
| $net_export_{n,y,t}$ | Net export of electricity [MWh] |
| $ntc_{n,nn}$ | Net transfer capacity from n to nn [MW] |
| $ntc_{nn,n}$ | Net transfer capacity from nn to n[MW] |
| $\eta_{s}$ | Charge to discharge efficiency [%] |
| $voll^{Load}$ | Load shedding costs [EUR/MWh] |

**Variables**

| | |
|---|---|
| $CAP_{n,p,y}$ | Capacity investment of conventional power plants [MW] |
| $CAP_{n,r,y}$ | Capacity investment of renewable energy sources [MW] |
| $CAP_{n,s,y}$ | Capacity investment of storage technologies [MW] |
| $CAP^{RTO}_{n,p,y,t}$ | Operational conventional generation capacity [MW] |
| $CAP^{Up}_{n,p,y,t}$ | Increase of conventional operational capacity [MW] |
| $CAP^{Down}_{n,p,y,t}$ | Decrease of conventional operational capacity [MW] |
| $CHARGE_{n,s,y,t}$ | Charge of storage [MWh] |
| $DISCHARGE_{n,s,y,t}$ | Discharge of storage [MWh] |
| $COST$ | Total system cost (objective value) [EUR] |
| $COST^{Gen}_{y}$ | Generation and curtailment costs [EUR] |
| $COST^{Inv}_{y}$ | Investment in generation and storage capacity [EUR] |

| | | | |
|---|---|---|---|
| $cvar_{p,y}$ | Variable generation costs for conventional power plants [EUR/MWh] | $CURT^{Gen}_{n,r^{curt},y,t}$ | Curtailment of renewable technologies [MWh] |
| $cvar_{r,y}$ | Variable generation costs for renewable energy sources [EUR/MWh] | $FLOW_{n,nn,y,t}$ | Export flow of node n to nn [MWh] |
| $demand_{n,y,t}$ | Electricity demand [MWh] | $FLOW_{nn,n,y,t}$ | Import flow of node nn to n [MWh] |
| $duration_s$ | Duration to charge the storage [h] | $GEN_{n,p,y,t}$ | Generation from conventional power plants [MWh] |
| $d2c_ratio_s$ | Discharge to charge ratio [%] | $GEN_{n,r,y,t}$ | Generation from renewable energy sources [MWh] |
| $gen_{n,hy,y,t}$ | Generation from hydropower plants [MWh] | $GEN^{Full}_{n,p,y,t}$ | Generation from conventionals at full load [MWh] |
| $gmin_p$ | Minimum generation level [%] | $GEN^{Min}_{n,p,y,t}$ | Generation from conventionals at minimum load [MWh] |
| $grid_losses$ | Grid losses [%] | $LEVEL_{n,s,y,t}$ | Storage level [MWh] |
| $lifetime^{eco}_{p,r,s}$ | Economic lifetime [y] | $SHED^{Load}_{n,y,t}$ | Load shedding [MWh] |

### 3.1.1 Objective Function

The objective of the model, as defined in Eq. (1), is to minimise the total system cost for each year *(y)*. Total system costs are summed investment costs, defined in Eq. (2), and generation costs, defined in Eq. (3). In Eq. (2), investment decisions for all technologies[2], i.e. $CAP_{n,p,y}$ for conventional power plants (index $p$), $CAP_{n,r,y}$ for renewable energy sources *(r)*, and $CAP_{n,s,y}$ for storages *(s)*, are optimised endogenously.[3] The set-up of the optimisation problem implies that all technologies with positive invested capacity cover their fixed costs according to [33]. Investment costs as well as fixed operation and maintenance costs (FOM) are combined in one parameter per class of technology, e.g. $cinv_{p,y}$ for conventional and $cinv_{r,y}$ for renewable generation technologies and $cinv_{s,y}$ for batteries, with the latter comprising both the cost for the batteries' power and energy storage. Investment costs are represented as annuities, and cost values are explained in 3.2.1. The generation costs are shown in Eq. (3). The model also considers cost for curtailment of non-dispatchable renewables and load shedding.

$$minCOST_y = COST^{Inv}_y + COST^{Gen}_y \qquad \forall y \qquad (1)$$

$$COST^{Inv}_y = \sum_n \left( \sum_{p,y} cinv_{p,y} * CAP_{n,p,y} + \sum_{r,y} cinv_{r,y} * CAP_{n,r,y} + \sum_{s,y} cinv_{s,y} * CAP_{n,s,y} \right) \qquad \forall y \qquad (2)$$

[2] Hydropower constitutes the exception and is represented through exogenous generation time series, whereas pumped-hydro storage is modelled with fixed installed capacities.

[3] The model has an index *n* for different nodes within a system, which allows us to model markets with nodal pricing.

$$COST_y^{Gen} = \sum_n \left( \sum_{p,t} cvar_{p,y} * GEN_{n,p,y,t} + cramp_{p,y} * \left(CAP_{n,p,y,t}^{UP} + CAP_{n,p,y,t}^{DOWN}\right) + \sum_{r,t} cvar_{r,y} * GEN_{n,r,y,t} + \sum_{r^{ndisp},t} ccurt_{r^{ndisp}}^{Gen} * CURT_{n,r^{ndisp},y,t}^{Gen} + \sum_t voll^{Load} * SHED_{n,y,t}^{Load} \right) \quad \forall y \quad (3)$$

### 3.1.2 Energy Balance Constraint

Eq. (4) guarantees that the electricity load *($load_{n,y,t}$)* is fully covered at every time step *(t)*. Power exchange with other systems, i.e. those beyond Germany or Texas, are considered exogenously in our model. The approach from [34] was taken to calculate the total system load as the sum of the realised electricity consumption *($demand_{n,y,t}$)* and net exports *($net_export_{n,y,t}$)* for each time step in each year. Furthermore, we assume the generation from non-dispatchable hydropower plants is exogenous and deduct it from demand. Consequently, $load_{n,y,t}$ reflects the load that must be covered with generation from available conventional and renewable capacities, storage dispatch, node interconnections (including grid losses *($grid_losses$)*), and load shedding if required.

$$load_{n,y,t} = demand_{n,y,t} + net_export_{n,y,t} - gen_{n,hy,y,t} = \sum_p GEN_{n,p,y,t} + \sum_r GEN_{n,r,y,t} + \sum_{rsv} GEN_{n,rsv,y,t} + \sum_s \left(DISCHARGE_{n,s,y,t} - CHARGE_{n,s,y,t}\right) + \sum_{nn} \left(1 - \frac{grid_losses}{2}\right) * FLOW_{nn,n,y,t} - \sum_{nn} \left(1 - \frac{grid_losses}{2}\right) * FLOW_{n,nn,y,t} + SHED_{n,y,t}^{Load} \quad \forall n,y,t \quad (4)$$

### 3.1.3 Electricity Generation from Conventional Power Plants

Eqs. (5), (6), (7), (8), and (9) describe the operation of conventional power plants. To capture partial-load efficiency and costs, Eq. (5) separates generation into two operational states: full-load output and minimum-load operation. This differentiation enables a representation of thermal power plant performance under varying system conditions. Eqs. (6) and (7) impose upper bounds via endogenous capacity investments and lower bounds via minimum generation percentage *($gmin_p$)* on generation output, conditional on the level of operational capacity. This operational capacity represents the portion of installed capacity committed for dispatch and is determined endogenously by Eq. (8). Eq. (9) constrains the operational generation capacity by the available share *($avail_{p,y,t}$)* of installed capacity. This availability accounts for planned maintenance services, which vary for different technologies.

$$GEN_{n,p,y,t} = GEN^{Full}_{n,p,y,t} + GEN^{Min}_{n,p,y,t} \qquad \forall n,p,y,t \qquad (5)$$

$$GEN^{Full}_{n,p,y,t} \leq CAP^{RTO}_{n,p,y,t} \qquad \forall n,p,y,t \qquad (6)$$

$$GEN^{Min}_{n,p,y,t} \geq gmin_p * \left(CAP^{RTO}_{n,p,y,t} - GEN^{Full}_{n,p,y,t}\right) \qquad \forall n,p,y,t \qquad (7)$$

$$CAP^{RTO}_{n,p,y,t} = CAP^{RTO}_{n,p,y,t-1} + CAP^{Up}_{n,p,y,t} - CAP^{Down}_{n,p,y,t} \qquad \forall n,p,y,t \qquad (8)$$

$$CAP^{RTO}_{n,p,y,t} \leq avail_{p,y,t} * CAP_{n,p,y} \qquad \forall n,p,y,t \qquad (9)$$

### 3.1.4 Electricity Generation from Renewable Energy Sources

Renewable electricity generation is categorised into dispatchable *($r^{disp}$)* and non-dispatchable technologies *($r^{ndisp}$)*. For dispatchable sources, such as bioenergy and geothermal power plants, generation is constrained by the available endogenous installed capacity, along with a yearly and spatial average availability factor *($avail_{rdisp,y,t}$)*, as specified in Eq. (10). Non-dispatchable sources may be curtailed, when necessary, and have an availability factor *($avail_{n,rndisp,y,t}$)* in hourly resolution for each node, as modelled in Eq. (11).

$$GEN_{n,r^{disp},y,t} \leq avail_{r^{disp},y} * CAP_{n,r^{disp},y} \qquad \forall n, r^{disp}, y, t \qquad (10)$$

$$GEN_{n,r^{ndisp},y,t} + CURT^{Gen}_{n,r^{ndisp},y,t} = avail_{n,r^{ndisp},y,t} * CAP_{n,r^{ndisp},y} \qquad \forall n, r^{ndisp}, y, t \qquad (11)$$

### 3.1.5 Modelling of Storage Systems

Storage restrictions are represented through Eqs. (12), (13), (14), (15), and (16). The level of charge is governed by Eq. (12) with a discharge efficiency factor *($\eta_s$)*, while Eq. (13) imposes an upper limit via the storage duration *($duration_s$)* on the storage level. To avoid a discharge in hour one, the level of any used storage technology is set to zero in the first hour in Eq. (14). Charging and discharging processes are constrained by the endogenous installed capacity (except pumped-hydro storage, where installed capacity is exogenous) and an annually average availability factor *($avail_s$)*, as specified in Eqs. (15) and (16). In Eq. (16) the loss from discharging to charging *($d2c_ratio_s$)* has also been considered, via a percentage ratio for each storage technology.

$$LEVEL_{n,s,y,t} = LEVEL_{n,s,y,t-1} + CHARGE_{n,s,y,t} * \eta_s - DISCHARGE_{n,s,y,t} * \frac{1}{\eta_s} \quad \forall n,s,y,t \quad (12)$$

$$LEVEL_{n,s,y,t} \leq CAP_{n,s,y} * duration_s \quad \forall n,s,y,t \quad (13)$$

$$LEVEL_{n,s,y,t=1} = 0 \quad \forall n,s,y,t \quad (14)$$

$$CHARGE_{n,s,y,t} \leq avail_s * CAP_{n,s,y} \quad \forall n,s,y,t \quad (15)$$

$$DISCHARGE_{n,s,y,t} \leq avail_s * CAP_{n,s,y} * d2c_ratio_s \quad \forall n,s,y,t \quad (16)$$

### 3.1.6 Modelling Electricity Exchange within the Systems

Electricity exchange within a system, i.e. between nodes, is determined endogenously. The maximal flow between two nodes is limited via aggregated net transfer capacities *($ntc_{n,nn}$ and $ntc_{nn,n}$)* in Eqs. (17) and (18).

$$FLOW_{n,nn,t} \leq ntc_{n,nn} \quad \forall n,nn,y,t \quad (17)$$

$$FLOW_{nn,n,t} \leq ntc_{nn,n} \quad \forall nn,n,y,t \quad (18)$$

## 3.2 Case Study: Data and Assumptions

We optimise subsidy-free investment in five renewable technologies (solar PV, onshore wind, offshore wind, bioenergy, and geothermal), five conventional technologies (nuclear, lignite, hard coal, CCGT, and OCGT), alongside battery storage in Germany and Texas. Installed capacities for hydropower, including run of river, reservoir, and pumped-hydro storage, are held constant at their 2024 levels. Each model run optimises one year with 8760 hours. To introduce the datasets, it is helpful to split the whole period of analysis into a historical and a future period. For the historical time period, we parametrise each year from 2015 to 2024. The future time period parametrises the years 2030, 2035, 2040, 2045, and 2050. The analysis for each year is greenfield because our aim is to analyse the specific situation during that particular year, normalising e.g. for legacy assets. The model purposefully ignores any policy targets on renewable expansion or emission reductions. Conventional power plant additions are not limited by fuel availability or political restrictions, such as e.g. Germany's phasing out of nuclear energy in the past and coal in the future.

Cost assumptions represent an important part in our analysis. We present and explain them in detail in section 3.2.1. Furthermore, the model contains other technology-specific parameters such as economic lifetime and efficiency for all technologies, carbon content, start-up costs and minimum generation limits, and annual availability. For storages, we keep the duration and discharge-to-charge ratio constant based on [35]. For pumped-hydro storages, we assume a storage duration of 6 hours and an annual availability of 97%; and for battery storages, 2 hours and 99%. In section 3.2.2, we define the renewable potentials and energy regions, with location-specific availability factor in hourly resolution. The availability factor for conventional and dispatchable power plants is based on an annual average from the literature (see Appendix A). For conventional technologies, we apply technology-specific availability factors that remain constant throughout the analysis. For instance, the available capacity for lignite is restricted to 87% of endogenously installed capacity across all hours to account for typical maintenance and outage rates. Finally, section 3.2.3 presents the assumptions regarding the electricity demands.

### 3.2.1 Cost Data

A vital input parameter in our analysis is the investment cost for all technologies, varying by year. Most literature sources do not provide investment cost data in annual granularity. In assessing investment cost, our analysis addressed this challenge by compiling an extensive literature survey, and deriving annual investment costs in two steps:

First, we set up a consolidated database of investment cost data for various years from the literature. In total, we used data from more than 10 literature sources [35–49], providing investment costs for electricity generation capacity. Investment cost comprises all annual fixed cost, i.e. also fixed cost for operation and management. Data covered either Germany or Texas and provided historical cost with observations between the years 2015 and 2023. We are not differentiating between systems, i.e. investment costs for Germany and Texas are assumed to be equal. Each data point was treated as the nominal cost in the respective year, expressed in euros. This process provides between 27 and 64 investment cost observations per technology.

Second, a trend curve is generated using a linear regression line from 2015 to 2024 in accordance with [50]. This smooths cost fluctuations between different years and data sources. The resulting annual values represent our investment cost assumption for each technology and year during the historical time period. A summary of the raw data for investment cost is shown in Appendix C. The resulting investment costs used as input in our energy system model for the historical years from 2015 to 2024 are shown in Appendix C. For the future time period, we linearly extrapolate the regression results on our historical cost values for solar PV, wind turbines, and batteries. As the capital costs for onshore wind have risen slightly in our period under consideration, study [51] is used as the methodological basis for estimating future costs. Here we use a linear reduction of 30% from 2024 costs to 2050. All other generation technologies are assumed to remain at their 2024 cost levels.

Next, we parametrise the technology-specific variable costs. For renewable technologies, variable costs (except bioenergy) are assumed to be zero throughout the analysis for both systems and all years. For the seven conventional technologies, variable costs consist of costs for fuel, different in both Texas and Germany. The model computes the variable costs for conventional technologies ($cvar_{p,y}$), as well as ramping costs for conventional generation ($cramp_{p,y}$). Furthermore, the cost for $CO_2$-emission allowances is added in Germany, where an emission trading system exists, but not in Texas. Variable costs vary between years. Our analysis assumes constant fuel prices within each year, set at the annual average of the various fuels (and $CO_2$ allowances in Germany). The specific values for all years (both historical and future) are provided in Appendix D.

### 3.2.2 RES Generation Potentials

Availability factors for RES vary not only between systems such as Germany or Texas, but also within these systems. An analysis from [52] for Germany demonstrates that the natural variability in solar and wind resource availability is strong, both temporally and across locations. While investors want to build RES units where the conditions are best, available space is limited, providing a natural constraint on available (land use) potential for RES. To address such space limitations within each system, we divide the onshore wind and solar generation potential of both systems into different regions.[4] Each region is specified by an installable RES potential (i.e. how many gigawatts of the specific RES technology can be installed per region) and a region-specific hourly availability factor, which varies by year from 2015 to 2024.

We start by clarifying the different regions: we divide Germany into four and Texas into three onshore wind regions. The specific classification for onshore wind is taken from [53] for Germany and [54] for Texas. Furthermore, we introduce four solar regions corresponding to global irradiation [55,56], visualised in Figure 1. For Germany, we define irradiation thresholds for the different regions, with solar region 1 (SR1): < 1121 kWh/m², SR2: 1121-1180 kWh/m², SR3: 1181-1220 kWh/m², and SR4: > 1220 kWh/m². For Texas, we apply irradiation thresholds for the different regions, with SR1: < 1900 kWh/m², SR2: 1901-2115 kWh/m², SR3: 2116-2260 kWh/m², and SR4: > 2260 kWh/m².

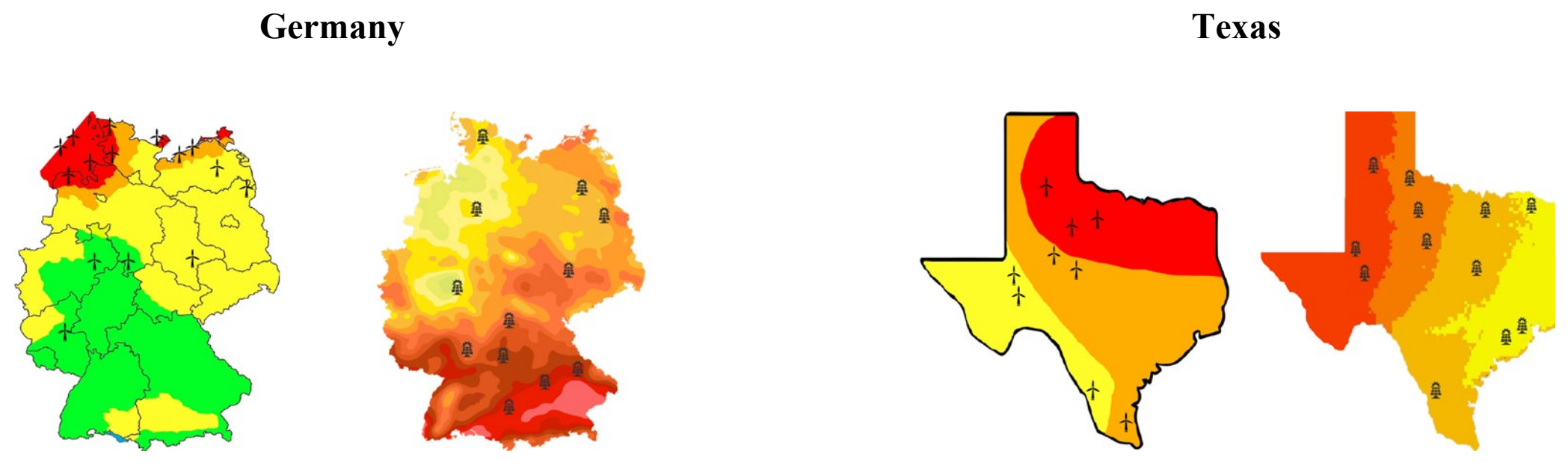

**Figure 1. Wind regions (left) and solar regions (right) for Germany and Texas**

As a next step, we specify the installable (land use) potentials for each onshore wind and solar region individually. From the definition of the different regions, we can compute how spacious each region is (in km$^2$), directly based on the maps in Figure 1. In the details, we use a visual clustering algorithm, which filters out background and artifact colours, and enables a quantitative analysis of colours' spatial share on the total system area. Once the size of each region has been determined, we need to take into account that only a fraction of the landmass in each region is available for RES installations. Based on the literature ([73,74]), we assume this fraction is 2% for onshore wind in both Germany and Texas and 2.5% for solar power, also identical in both systems. Finally, this number is divided by the required space per MW of installed capacity for the two technologies (4.1 ha/MW for onshore wind in both systems, 1.05 ha/MW for solar PV in both systems). These numbers are taken from [57,58] for onshore wind and for solar PV. This results in an upper limit of installable MW per region. Table 1 shows the individual values of the onshore wind regions (WR) and solar regions (SR). We assume these upper installation limits in our historical and future time periods.

[4] Bioenergy and geothermal technologies are modelled using annually constant, technology-specific availability factors.

**Table 1. Calculation of upper capacity potentials per onshore wind and solar region**

| | **Germany** | | | | **Texas** | | | |
|---|---|---|---|---|---|---|---|---|
| Region size (ha) | 35 759 200 | | | | 69 562 100 | | | |
| Onshore wind | WR1 | WR2 | WR3 | WR4 | WR1 | WR2 | WR3 | WR4 |
| Share of region (%) | 39 | 50 | 6 | 5 | 21 | 45 | 34 | - |
| Usable area (%) [59] | 2 | | | | 2 | | | |
| Space requirement (ha/MW) [57] | 4.1 | | | | 4.1 | | | |
| Upper limit (GW) | 68 | 87 | 10 | 9 | 71 | 153 | 115 | - |
| Solar | SR1 | SR2 | SR3 | SR4 | SR1 | SR2 | SR3 | SR4 |
| Share of region (%) | 29 | 39 | 20 | 12 | 20 | 33 | 18 | 29 |
| Usable area (%) [60] | 2.5 | | | | 2.5 | | | |
| Space requirement (ha/MW) [58] | 1.05 | | | | 1.05 | | | |
| Upper limit (GW) | 247 | 332 | 170 | 102 | 332 | 547 | 298 | 481 |

To obtain the region-specific hourly availability factors for each year of the analysis, we take the locations of the three largest solar PV and onshore wind power parks in each region.[5] The data for Germany are based on [61–63] and for Texas are based on [64–66]. Figure 1 depicts the location of the three largest sites. [67] provides hourly wind speeds and solar irradiation for all years of our historical time period. We use them to calculate three availability factors per region. Furthermore, to avoid overestimating the availability factors of wind turbines, the bias correction factor for MERRA-2 data from [68] is used for both systems. We use the unweighted hourly average of these three values as the hourly availability factor for that region during the historical time period of our analysis. For the availability factors in our future time period, we develop a reference weather year for each individual solar, on-, and offshore wind region for Germany and Texas from the annual averages between 1980 and 2024, based on the availability of data from [67]. The median year of the annual averages is used with the associated hourly availability factors for each region. The median years can be found in Appendix E. We therefore use the same weather conditions for each year in our future analysis. This approach preserves real time series of availability factors and prevents individual technologies from being favoured by only taking a single year.

[5] The availability factors for offshore wind in Germany are also determined with the three largest parks. As Texas does not currently have any offshore wind farms installed, we refer to the three largest planned projects.

Generation from hydropower plants, including run of river and reservoir, as well as pumped-hydro storages, are incorporated exogenously in all years at their observed 2024 capacity levels for Germany and Texas [69]. No endogenous capacity expansion is permitted for these technologies, reflecting their limited future expansion potential. Consequently, hydropower is modelled as a baseload generation technology. The generation is represented through exogenous hourly generation time series ($gen_{n,hy,y,t}$), which are derived from historical data. The corresponding hydropower generation time series are adopted from [73] for Germany and from [71] for Texas. Installed capacities for pumped-hydro storage are taken from the same respective sources. According to [72], Texas has no run-of-river plants or pumped-hydro storages in its system.

### 3.2.3 Demand and Net Exports

In order to model national electricity interchange, we take into account the different electricity pricing systems. Germany has a zonal pricing system, whereby the entire system acts as a single price node in the European day-ahead market. Therefore, Germany is modelled as a single-node system ($n = 1$), meaning that no national interchange is considered. For Texas, the electricity network of the largest transmission system operator ERCOT is used. ERCOT has a nodal pricing system with over 17,000 nodes, of which 900 have a unique locational marginal price and are labelled as settlement points. We consider Texas as a simplified seven-node system ($n = 7$) based on the defined ERCOT load zone of [74]. Each ERCOT load zone represents a single node in our model – except for the North Central and East zones, which are merged into a single zone due to the missing net transfer capacity from [75].

For the historical time period, hourly electricity demand and net exports are available from [76] for Germany. For each node in Texas, demand data are provided by [77], while net exports are obtained from [71]. Hourly net demand is simply calculated as the exogenous electricity demand added to the net exports. For clarification, only cross-border flow is considered here, as national electricity interchange is modelled endogenously by the maximum values out of [75], according to section 3.1.6. We do not consider transmission line expansion in our study, and set the transfer capacities as fixed over our entire observation period. With regard to our future time period, we assume a 100% increase in electricity consumption for both Germany (based on [78]) and Texas (based on [79]) by 2050. We use 2024 as a reference year, i.e. 2050 electricity demand in Germany in hour t is calculated as twice the electricity demand in that hour in the year 2024. Intermediate years (2030, 2035, 2040, 2045) in our future time period are interpolated between the 2024 and the 2050 values.

## 3.3 Limitations

One limitation of our study lies in our greenfield approach, which allows endogenous capacity expansion in annual resolution. Due to this approach, we focus purely on the situation in the year in question and do not consider the time periods before and after. While this is common in the literature – all optimisations focusing on annuities implicitly assume the system stays constant over the whole life-time of the investments – this abstracts from real-world dynamics. For example, a comparison between consecutive years can show unrealistic capacity changes. At the same time, this is a feature of our analysis as it allows us to assess the profitability of renewable technologies under the specific situation of any given year of observation.

A key driver for increasing RES competitiveness in our historical analysis is the reduction in the investment cost for renewable energy that was historically observed. However, this cost development is often attributed to learning curves, i.e. the result of installing large amounts of renewable capacity. Hence, it is uncertain to what extent these cost reductions would have manifested without subsidies, i.e.

to what extent RES would have become profitable if much less capacity would have been added. The dynamic effects of learning are left for future work.

Furthermore, the energy system model presented in section 3.1 should be mentioned, as it strikes a balance between the complexity needed to properly answer the empirical research question on the one hand and keeping the model short and intuitive, as well as the outputs explainable, on the other hand. As a result of this trade-off, model-specific limitations remain that should be kept in mind when interpreting the results. First, some of the features discussed in other energy system modelling publications are not considered, for example uncertainty and endogenous sector coupling (see e.g. [80]), or endogenous cross-border electricity exchange. The latter is considered exogenously via hourly historical data series. While this preserves the exchange pattern of each year, it prevents the model from representing how trade would adjust in a competitive equilibrium. Second, our analysis of the future time period depends on the assumptions regarding fixed costs and variable costs of different technologies until 2050. These are inherently difficult to forecast. For fixed costs, we assume that the decreasing trend in investment cost for RES will continue, while we assume that conventional technologies remain at their 2024 levels. This may favour the expansion of renewables, as the cost developments of the other technologies may also be subject to learning effects and cost reductions. For variable costs, we assume that fuel prices and $CO_2$ prices develop as discussed in section 3.2.1. If fuel and $CO_2$ prices were to be higher, renewables would be more profitable; if they turn out to be lower, shares of RES in the optimal systems would be lower. All in all, future results react strongly to this uncertainty, and other assumptions regarding these parameters will lead to other results. Consequently, our future analysis is not meant as a forecast but more of an if-then analysis.

# 4 RESULTS AND DISCUSSION

This section presents the results for Germany and Texas (ERCOT). First, we focus on the analysis of the historical cost-optimal expansion of energy technologies in annual resolution from 2015 to 2024. Second, we examine the evolution of the pathway through 2050 based on the further development of parameters from the previous years. Third, we determine the effects of six influencing factors using a ceteris paribus approach.

## 4.1 Historical Cost-Optimal Expansion

Figure 2 depicts the annual share of technologies in total generation for the German (left) and ERCOT (right) electricity systems.

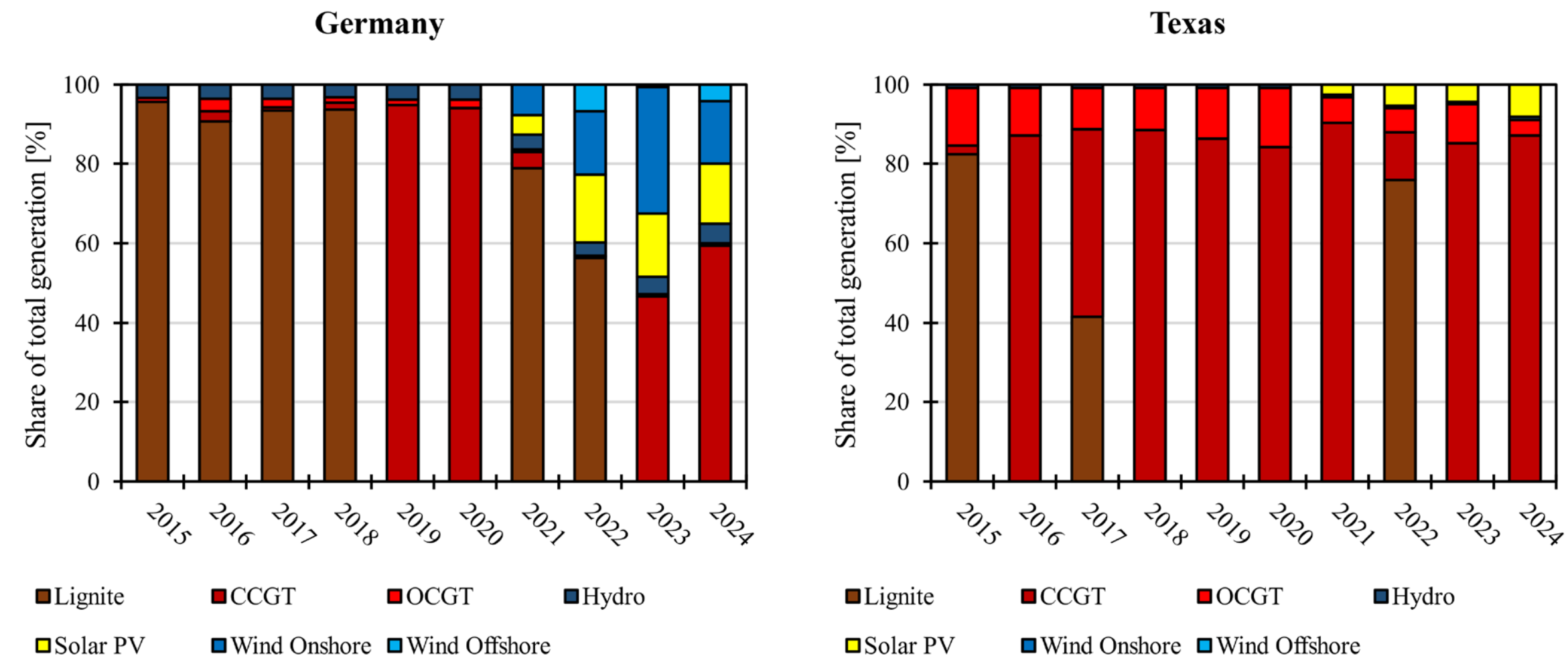

**Figure 2. Historical time period with annual share of total generation from 2015 to 2024**

Three fundamental aspects can be derived from the figure. First, renewable generation starts to become competitive in both systems after the year 2020, rising to 35% in Germany in the year 2024. The highest RES share comes from onshore wind, with PV being nearly on par for most years. The RES share in Texas lies below 10% in all years, but with an increasing trend over time for solar PV.

Second, when compared to reality, our results are 11 percentage points lower than the real observations in Germany for the year 2024 [76]. Texas has a difference of 28 percentage points [81]. Due to the lack of support mechanisms in our model, it was to be expected that modelled values are lower than real world observations.

Third, the majority of the electricity in both systems is provided by conventional technologies. The model favours lignite and gas-fired power plants over hard coal. There is no investment in nuclear, bioenergy, and geothermal capacity.

For a more detailed analysis, we look at the period from 2015 to 2020 first, where the combination of cheap fuel prices, high investment costs for RES, and a low carbon price make conventional generation so competitive that it covers 100% of demand. In reality, financial support schemes pushed the share of solar PV and wind to around 20% in Germany (based on [70]) and 12% in Texas in 2015 ([77]). The year 2021 marked the beginning of a second period, wherein a RES share of almost 12.5% for Germany and 2.5% for Texas is already part of the optimal system, purely on a cost basis. While these results are below reality, it demonstrates RES being close to competitiveness even at moderate gas and $CO_2$ prices, especially in Germany. From 2021 to 2024, the share of electricity generation from solar PV, onshore wind, and offshore wind grew further. The peak in the year 2023 reached more than 50% in Germany, driven by the turmoil conventional supply was facing after the Russian attack on Ukraine. This development can also be attributed to a sharp increase in the carbon price in 2023, which more than doubled compared to 2020 [82]. In Texas, even with more favourable environmental conditions for renewables, the renewable share is considerably lower compared to Germany, with conventional electricity generation still accounting for more than 90% of all electricity generation in 2024.

## 4.2 Future Cost-Optimal Expansion

Figure 3 shows the annual generation shares for the years from 2030 to 2050 in 5-year steps, with model results for the year 2024 for comparison.

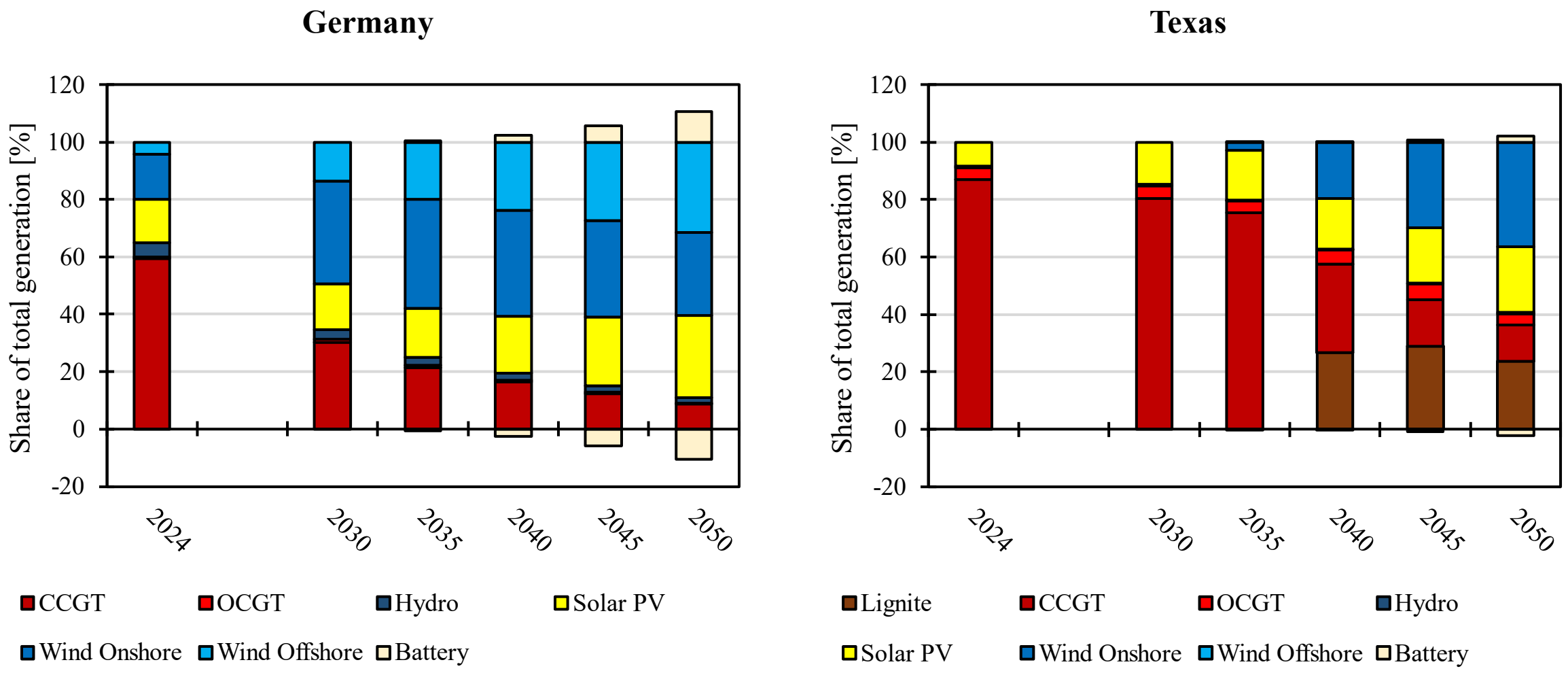

**Figure 3. Future period with annual share of total generation from 2030 to 2050 in 5-year steps, with 2024 as the last historical year**

Our results show a steady increase in renewable generation in both systems up to 2050. The fast increase of the German RES share, going from 35% in 2024 to 89% in 2050, correlates with a strong rise in the $CO_2$ price. Having more than 80% of total generation from RES by 2040 even without any subsidy is a vital sign of how competitive RES can become in the future European energy system. Nonetheless, our model does not reach the policy target of a carbon-neutral German system by 2045. The increase in RES generation is less pronounced in Texas, driven by the assumed absence of a $CO_2$ price for the whole period of observation. Nonetheless, the expected decline in investment costs of RES and the expected rise in fuel costs drive a steady growth in Texas as well. The RES share rises from 8% in 2024 to 59% in 2050. In both systems, renewables complemented with gas-fired power plants form the cost-optimal future generation system. In Texas, lignite-fired power plants become an additional part of the electricity mix after 2040 due to comparably low fuel prices and the lack of carbon pricing.

In the details, there are additional differences between Germany and Texas in RES expansion. From the outset, Germany relies more on wind energy, both from on- and offshore. Wind increases by 13 (onshore) and 27 percentage points (offshore) from 2024 to 2050. The share of solar PV increases by 13 percentage points, from 15% to 28% percent of total electricity generation. This difference shows that the favourable wind conditions, particularly in the North and Baltic Seas, as well as in the coastal regions, offer a great potential if investment costs continue falling.

In Texas, solar PV is the main renewable technology until 2035, when onshore wind also becomes competitive. The share of solar PV increases from 8% in 2024 to 23% in 2050, even without an established emission trading system. The expansion of onshore wind starts in 2035 with a share of 3% and increases to 36% in 2050. This underscores that Texas has a high potential for onshore wind in addition to solar power.[6] The expected decline in investment costs for RES thus correlates with a steadily increasing generation share in Texas. Figure 4 illustrates the overall development of the RES share in both systems.

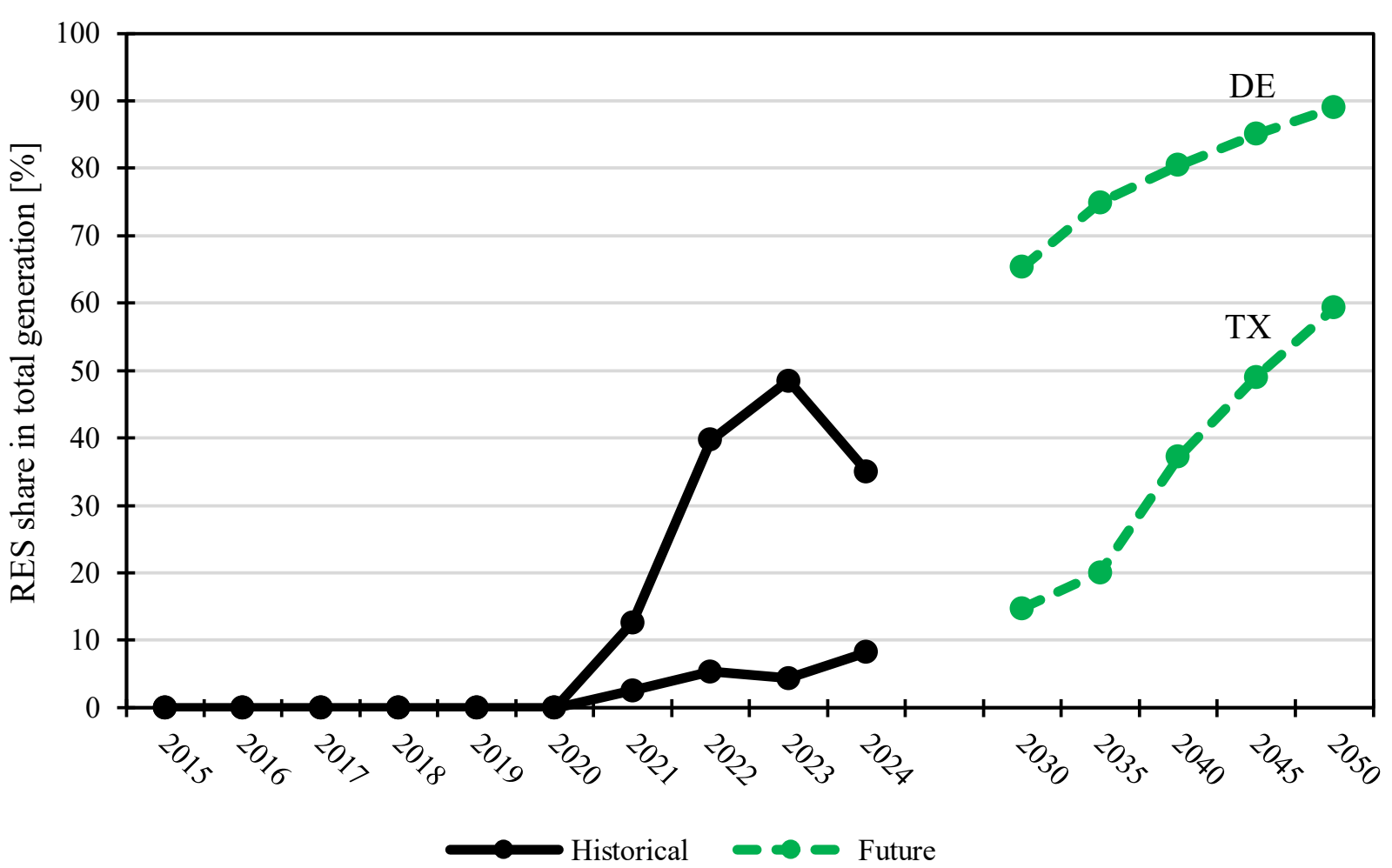

**Figure 4. RES share in total annual generation from 2015 to 2024, followed in 5-year steps from 2030 to 2050 for Germany and Texas**

The development of the RES share in total generation highlights two further facts. First, regardless of the system, 2021 emerges as a turning point for RES expansion. This initial share of 13% in Germany and 3% in Texas has steadily increased in both and thus represents a fixed component in the future electricity system. With regard to Germany's political goal of an 80% share of RES by 2030, our purely market-driven share amounts to approximately 65%, which once again underlines the rising RES competitiveness in, and positive outlook for, the future. Texas does not have political goals in this regard. Second, there are two different paths for both systems in the future. While Texas' share of RES is rising steadily due to the falling investment costs, the curve for Germany is much higher but gradually flattening out despite an additional rise in the $CO_2$ price.

[6] In contrast, offshore wind will not even in the long term become part of the cost-optimal system, which is justified by the comparably low offshore wind speeds in the Gulf of Mexico (see Appendix F).

## 4.3 Identification of Driving Factors

In both systems, the cost-optimal share of generation from RES is 0% in 2015 (see Figure 2). This percentage increases to 35% in Germany and 8% in Texas in 2024. While this year also reflects the peak value in Texas, the highest share in Germany is around 48% in 2023. We argue in section 4.1 that falling investment costs for RES and rising prices for fuel and $CO_2$ emissions (the latter especially in Germany) correlate with improvement in RES competitiveness. However, correlation is not necessarily causality.

Hence, we provide the following analysis that quantifies the individual effects of each parameter change on the RES share of total generation *($\Delta s_j^{RES}$)*, holding everything else constant (ceteris paribus). On this basis we determine the influence of six different factors *(j)*: weather, carbon price, fuel costs, investment costs for RES, demand, and cross-border flow. We insert the values from 2024 individually into the fundamental model from 2015 and analyse the impact based on [83]. We thus calculate a harmonised change of RES share in total generation brought about by the respective factor, defined in Eq. (19).

$$\Delta s_j^{RES} = \frac{GEN_j^{RES}}{GEN_j^{RES} + GEN_j^{CONV}} - \frac{GEN_{2015}^{RES}}{GEN_{2015}^{RES} + GEN_{2015}^{CONV}} \quad (19)$$

This approach assumes a simplified, linear interaction between the different parameters. This means that the sum of the six individual RES shares does not necessarily add up to the total share from 2024 *($s_{2024}^{RES}$)* according to Eq. (20). Non-linear interactions, which are often present in real-world situations, are approximated as being responsible for the remaining difference *($\varepsilon$)* in Eq. (21).

$$s_{2024}^{RES} \neq \sum_{j=1}^{6} \Delta s_j^{RES} \quad (20)$$

$$\varepsilon = s_{2024}^{RES} - \sum_{j=1}^{6} \Delta s_j^{RES} \quad (21)$$

The changes are interpreted as the contribution of each respective factor to the total change in RES shares from 2015 to 2024. The reader is also referred to Appendix G and Appendix H, which contain a sensitivity analysis for all parameters from this analysis, and an analysis whereby the installation of conventional capacities is limited to their historical maximum. Figure 5 summarises the individual contributions of all six factors and shows the harmonised change in percentages of RES in total generation between 2015 and 2024.

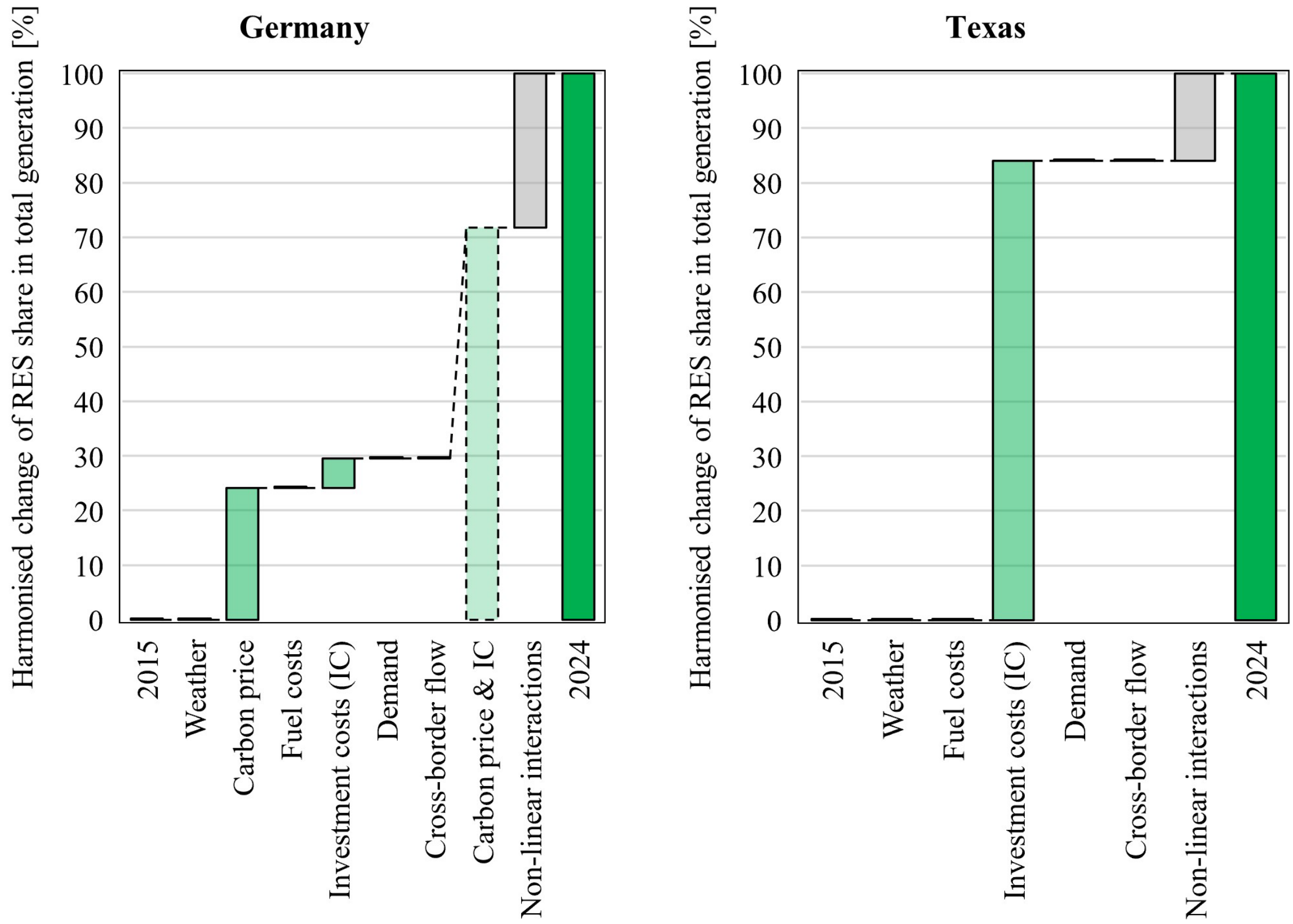

**Figure 5. Harmonised change of RES share in total generation in 2015, showing the contribution of individual factors and additional sum of carbon price and investment costs for Germany and for Texas**

The figure reveals three major findings. First, the carbon price is the most important factor for the rising RES share in Germany, reaching 24% of the total increase. This shows how vital the EU ETS is in driving the energy transition. The absence of this mechanism in Texas can therefore be considered a reason for the overall lower RES share in the cost-optimal system. The development of investment costs has the second-largest effect in Germany and the largest in Texas. The investment cost variation alone reaches approximately 6% of the total change in Germany, and 84% in Texas. The higher proportion in Texas is again attributable to Texas not having a $CO_2$ price, which reduces the total change and eliminates this factor as a driver for change at the same time.

Second, our results show that some factors have no linear effect (0%) on the RES share in either system. These are inter-year variations in the weather (including the capacity factors of solar and wind), in the fuel costs, in the electricity demand, and in the cross-border flow. This is surprising for the fuel price of natural gas, especially in Germany, which increased in the context of the Russian attack on Ukraine. However, the gas price had already declined somewhat from its peak in 2023. Furthermore, it needs to be repeated that a factor not showing a linear effect can still contribute to a non-linear effect – similar to the additional analysis we presented for a combined change of $CO_2$ price and investment cost in Germany.

Third, the non-linear interaction term is comparably large in Germany, implying that the sum of the six individual factors does not equal the observed increase in RES shares from 2015 to 2024 when all six factors are varied simultaneously. Observing the high share of the non-linear interactions in Germany, we additionally show the combined contribution of the $CO_2$ price and the investment cost changes. Varying these two factors simultaneously already reaches about 72% of the overall effect, while the non-linear interaction share decreases by 42% (as the remaining four factors still have an effect of zero). This indicates that the increase of the German RES share can mostly be attributed to the joint influence of carbon pricing and investment cost development.

# 5 CONCLUSION

A key motivation for this study is the ongoing debate regarding the competitiveness of RES. Neither rising RES shares nor simple cost metrics such as LCOEs can solve this debate. Even after more than 25 years of energy transition, the question as to what extent renewable energy is competitive remains unanswered – and controversial. This is the research question our study answers. Methodologically, this study presents a partial equilibrium energy system model, which optimises investment and dispatch of generation technologies to minimise total system cost. Empirically, we parametrise the model to analyse the cost-optimal integration of renewable energy technologies into two different electricity systems: Germany and Texas. Our research

- establishes a modelling framework that can be used to analyse cost-efficient RES shares in any electricity system;
- quantifies a counter-factual, cost-optimal expansion of renewables in Germany and Texas for a period from 2015 to 2024, as well as for 2030, 2035, 2040, 2045, and 2050;
- and quantifies the individual impact of six different factors on increasing RES shares.

Our systemic perspective can thus answer the question as to what extent RES have been competitive since 2015. We specifically analyse the situation during individual years, always focusing on annualised investment cost. The key insight in our study is that RES are increasingly competitive.

As of today, a significant share of RES would be present in both systems even without subsidies, mostly driven by declining investment costs in RES technologies, which improved their profitability. The share is significantly higher in Germany (35% of total generation in 2024) than in Texas (8%) – despite availability factors for both solar and onshore wind being better in Texas. The difference is mostly driven by the $CO_2$ price in Germany, which increases the costs of conventional technologies. For the medium- to long-term future, our results show a further increase in the share of RES, reaching 89% in Germany and 59% in Texas by the year 2050. This future outlook, however, depends on key assumptions such as e.g. a further increase of $CO_2$ prices in Europe and an extrapolation of RES investment cost declines.

However, while confirming the increasing competitiveness of RES, these results are below real-world shares, both in historical observation (Germany and Texas) and for future plans (e.g. Germany aiming for national climate neutrality in 2045). Nonetheless, we can conclude that renewable energy can be expected to grow even when subsidies are reduced significantly, but not at the same rate. Our results suggest that especially the last 10% of decarbonisation is hard to achieve on a purely market basis and will thus require high subsidies to achieve.

Once the increasing competitiveness of RES has been established, an interesting follow-up question is how much each factor contributed to this increase. We consider carbon cost, investment cost for RES, fuel costs, weather, demand, and cross border flows and determine their respective individual contribution in a ceteris paribus analysis. The analysis reveals that falling investment costs for RES in both systems, coupled with the $CO_2$ price in Germany, contributed most to the increase in RES shares from 2015 to 2024.

Our work is insightful for policy makers aiming to adapt future subsidy frameworks, investors seeking to estimate the costs of renewable energy systems, and critics concerned about the high levels of subsidisation and fluctuating availability of renewable energy sources. This helps policymakers focus their support where it is actually needed and avoid ineffective measures. For example, as $CO_2$ emissions have negative externalities, not pricing $CO_2$ can be interpreted as an implicit subsidy for conventional generation. Internalising these external effects, as the European emission trading system does, proves a very effective tool in our analysis for promoting the expansion of renewable energy and driving the energy transition forward.

Furthermore, investors gain a clearer overview of the rising competitiveness of solar PV and wind energy in the two different electricity systems of Germany and Texas. At the same time, our analysis helps to put concerns about fluctuating renewable generation into perspective by showing how a purely market-based system could look. Overall, the findings can support more targeted decisions and contribute to a more robust and affordable energy transition.

# APPENDIX

## Appendix A. Technical Parameters with Data Sources

Table 2 shows the associated data sources for the input parameters of the greenfield investment model.

**Table 2. Technical parameters with data sources**

| Description | Parameter | Source |
|---|---|---|
| **System-specific** | | |
| Hourly electricity consumption | $demand_{n,y,t}$ | DE: [76], TX: [77] |
| Hourly cross-border flow | $net_export_{n,y,t}$ | DE:[76], TX: [71] |
| Curtailment costs for non-dispatchable technologies | $ccurt_{r(ndisp),y}^{GEN}$ | DE, TX: [84] |
| Value of lost load | $voll^{Load}$ | DE, TX: [32] |
| Grid losses | $grid_losses$ | |
| **Technology-specific** | | |
| Hourly availability factors for non-dispatchable technologies | $avail_{n,rndisp,y,t}$ | DE, TX: [67] |
| Hourly hydropower production time series & installed pumped-hydro capacities | $gen_{n,hy,y,t}$ | DE: [73], TX: [71] |
| Yearly availability factors for dispatchable RES technologies | $avail_{rdisp,y,t}$ | DE, TX: [35] |
| Yearly availability factors for conventional power plants | $avail_{p,y,t}$ | |
| Minimum generation level for conventional power plants | $gmin_p$ | |
| Yearly availability factors for storage technologies | $avail_s$ | DE, TX: [32] |
| Discharge efficiency factor | $\eta_s$ | |
| Storage duration | $duration_s$ | |
| Discharge to charge ratio for storage technologies | $d2c_ratio_s$ | |
| **Further data for calculating parameters within the model** | | |
| Annualised Investment cost for conventional power plants ($cinv_{p,y}$), renewable energy sources ($cinv_{r,y}$), and battery storage systems ($cinv_{s,y}$) are determined in the model using the following inputs: | | |
| Interest rate | | DE, TX: [85] |

| Economical lifetime | DE, TX: [35] |
|---|---|
| Investment costs | see Appendix B and Appendix C |
| Fixed operation and maintenance costs | |
| Variable costs for renewable energy sources *($cvar_{r,y}$)* and conventional technologies *($cvar_{p,y}$)* are determined in the model using the following inputs: | |
| Efficiency | DE, TX: [35] |
| Carbon content | DE, TX: [86] |
| Fuel prices | see Appendix D |
| $CO_2$ price | |
| Ramping costs for conventional generation *($cramp_{p,y}$)* are determined in the model using the following inputs: | |
| Start-up fuel consumption | DE, TX: [35] |
| Start-up fix costs | |
| Fuel prices | see Appendix D |

# Appendix B. Investment Cost Graphs

The underlying respective sources for the regression are referenced in Table 3.

**Table 3. Respective sources for regression**

| | 2015 | 2016 | 2017 | 2018 | 2019 | 2020 | 2021 | 2022 | 2023 | 2024 |
|---|---|---|---|---|---|---|---|---|---|---|
| Nuclear | | | | | | | | | | |
| REF | [35,36,43,46,48,87] | [46,48,87] | [46,48,87] | [42,46,48,87] | [42,46,48,87] | [35,36,42,44,46,48,87] | [42,46,48,87] | [42,46,48,87] | [42,46,48,49,87] | - |
| Lignite | | | | | | | | | | |
| REF | [35,43,47,48] | [4,47,48] | [42,47,48] | [4,42,47,48] | [42,48] | [35,42,44,48] | [4,42,48] | [42,48] | [42,48] | - |
| Hard coal | | | | | | | | | | |
| REF | [35,37,46–48] | [36,46–49] | [42,46–48] | [4,42,46–48] | [42,46,48] | [35–37,42,46,48,49] | [4,42,46,48] | [42,46,48] | [42,46,48] | - |
| CCGT | | | | | | | | | | |
| REF | [35–37,43,46–48] | [46–49] | [42,46–48] | [4,42,46–48] | [42,46,48] | [35–37,42,44,46,48,49] | [4,42,46,48] | [42,46,48] | [42,46,48] | - |
| OCGT | | | | | | | | | | |
| REF | [35–37,43,46–48] | [46–49] | [42,46–48] | [4,42,46–48] | [42,46,48] | [35–37,42,44,46,48,49] | [4,42,46,48] | [42,46,48] | [42,46,48] | - |
| Solar PV | | | | | | | | | | |
| REF | [35–37,43,46,48] | [46,48,49] | [46,48] | [4,46,48] | [46,48] | [35–37,44,46,48,49] | [4,46,48] | [46,48] | [46,48] | - |
| Wind Onshore | | | | | | | | | | |
| REF | [35–37,43,46–48] | [46–49] | [37,42,46–48] | [4,37,42,46–48] | [37,42,46,48] | [35–37,42,44,46,48,49] | [4,37,42,46,48] | [42,46,48] | [42,46,48] | - |
| Wind Offshore | | | | | | | | | | |
| REF | [35–37,43,46,48] | [37,46,48] | [37,46,48] | [4,37,46,48] | [42,46,48] | [35–37,42,44,46,48,49] | [4,42,46,48] | [42,46,48] | [42,46,48] | - |
| Bioenergy | | | | | | | | | | |

| REF | [35–37,43,46–48] | [46–49] | [42,46–48] | [39,42,47,48] | [42,48] | [35–37,42,44,48,49] | [4,42,48] | [42,48] | [42,48] | - |
|---|---|---|---|---|---|---|---|---|---|---|
| Geothermal | | | | | | | | | | |
| REF | [35,36,43,46–48] | [46–48] | [46–48] | [46–48] | [46,48] | [35–37,44,46,48,49] | [46,48] | [46,48] | [46,48] | - |
| Battery Storage | | | | | | | | | | |
| REF | [4,36,40,43,45,48] | [40,45,48,49] | [40,42,45,48] | [40,42,45,48] | [40,42,45,48] | [36,40,42,44,48,49] | [4,42,48] | [38,40,42] | [42,48] | - |

Figure 6 visualises the cost data input for conventional technologies, renewable energy sources, and batteries. The raw data for each year, as gathered from the literature (see Table 3), is shown in the box plots. The black line results from a linear regression, used to smooth out fluctuations between years.

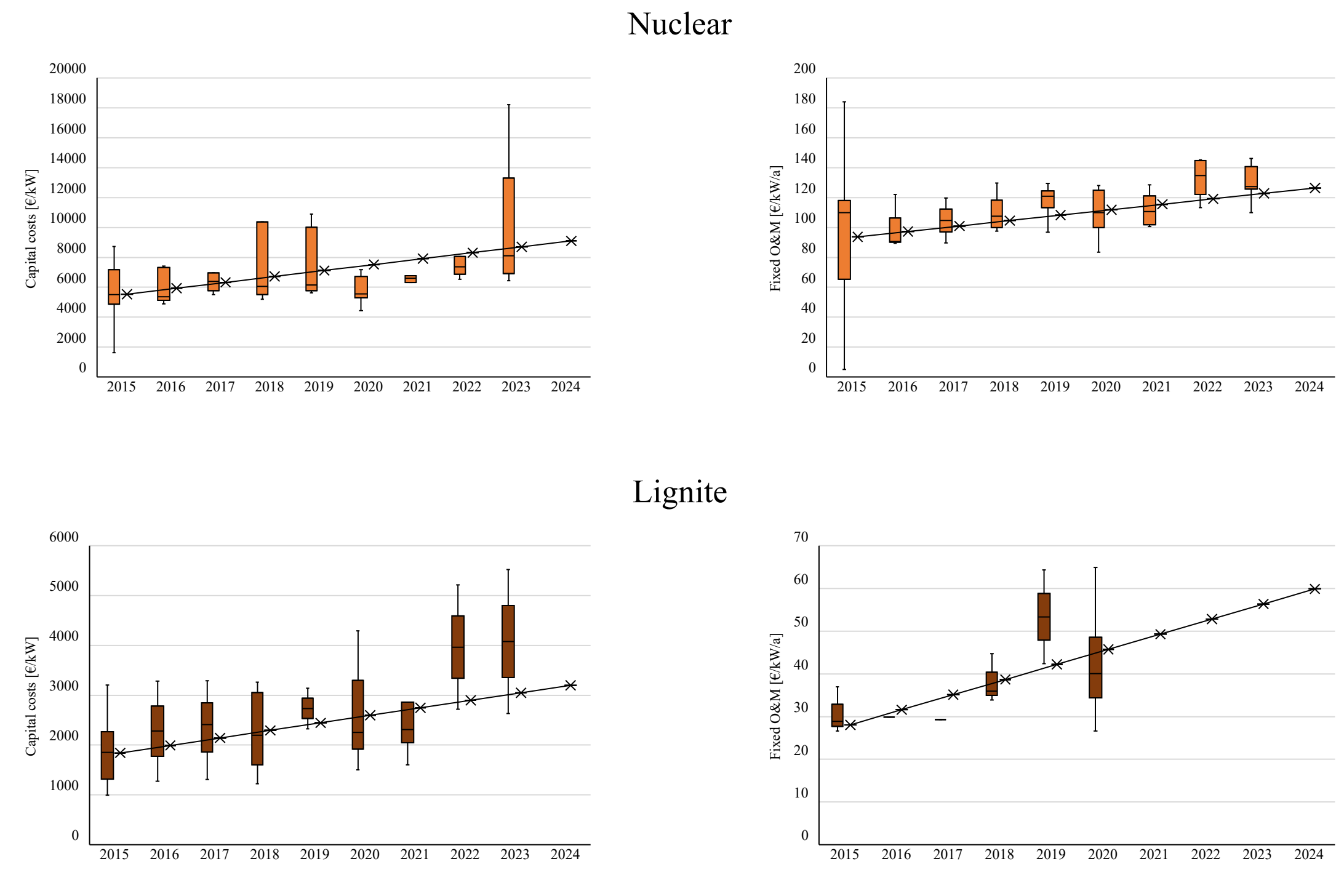

Hard Coal

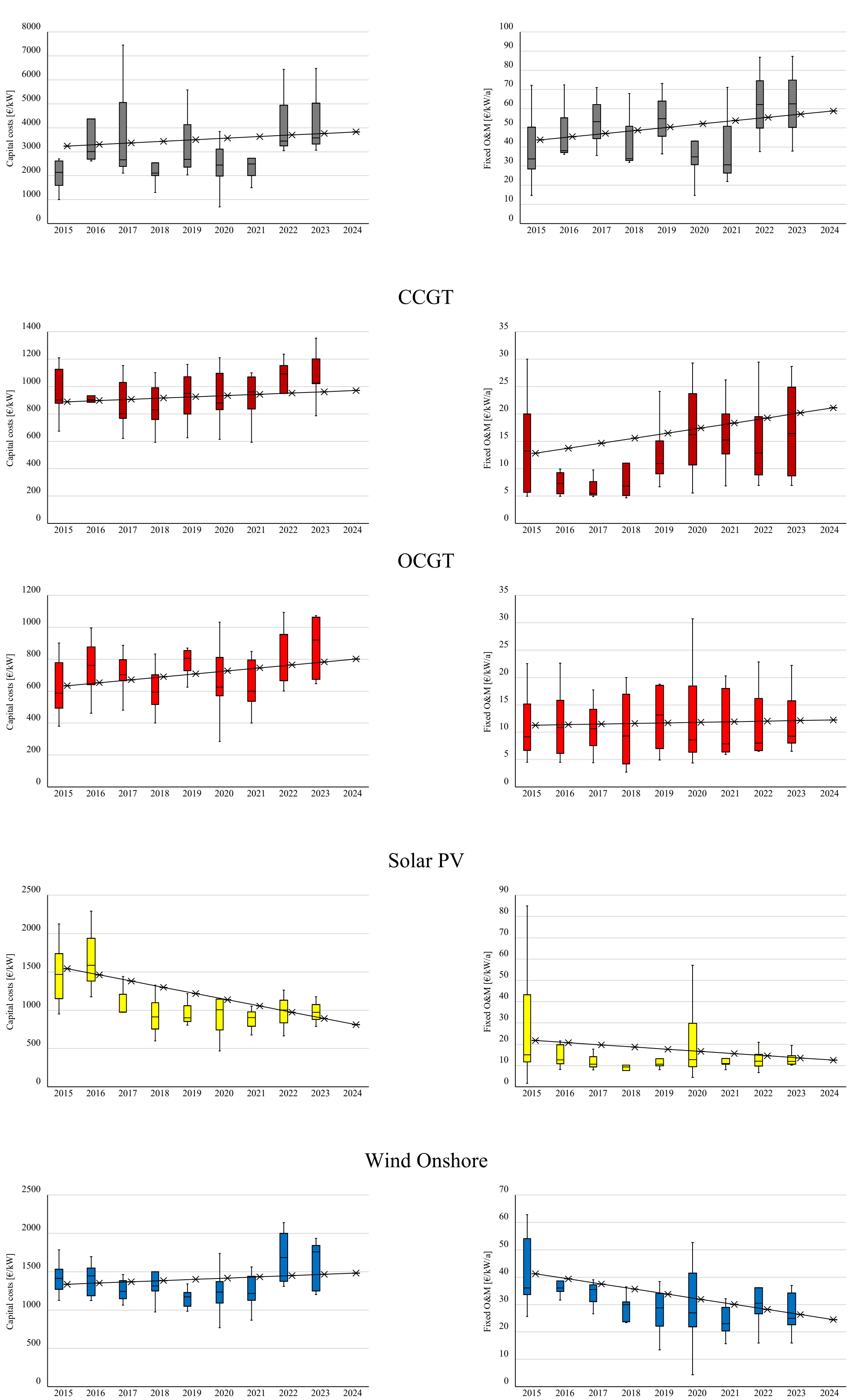

Wind Offshore

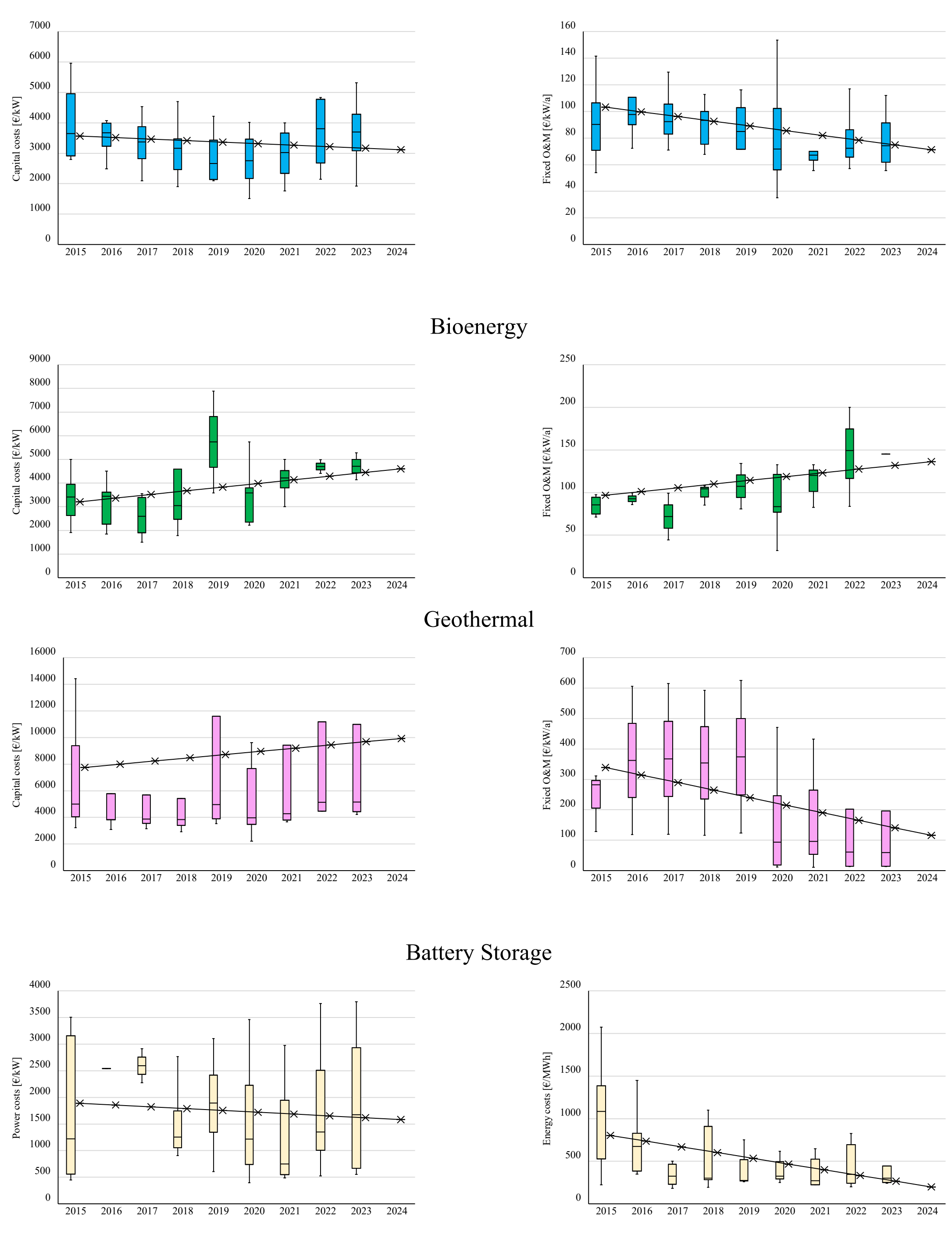

**Figure 6. Capital costs in €/kW (left) and FOM costs per year in €/kW/a (right) for each generation technology in annual resolution with linear regression line**

## Appendix C. Investment Cost Data

The cost structures of the investigated technologies are detailed in Table 4. We report both the fixed investment cost (IC) in €/kW and the FOM as annualised cost (€/kW/a). Given the annualised set-up of the optimisation model, we provide the annualised investment costs (AIC) as the final parameter ($cinv_{p,r,s,y}$), representing the sum of the input of annualised IC and FOM.

**Table 4. Historical cost data with source reference**

| | 2015 | 2016 | 2017 | 2018 | 2019 | 2020 | 2021 | 2022 | 2023 | 2024 |
|---|---|---|---|---|---|---|---|---|---|---|
| Nuclear | | | | | | | | | | |
| IC (€/kW) | 5539 | 5935 | 6332 | 6728 | 7124 | 7520 | 7916 | 8312 | 8708 | 9104 |
| FOM (€/kW/a) | 94 | 97 | 101 | 105 | 108 | 112 | 116 | 119 | 123 | 126 |
| AIC (€/kW/a) | 487 | 519 | 551 | 583 | 614 | 646 | 678 | 710 | 742 | 773 |
| Lignite | | | | | | | | | | |
| IC (€/kW) | 1841 | 1992 | 2143 | 2294 | 2445 | 2596 | 2746 | 2897 | 3048 | 3199 |
| FOM (€/kW/a) | 28 | 32 | 35 | 39 | 42 | 46 | 49 | 53 | 56 | 60 |
| AIC (€/kW/a) | 165 | 180 | 195 | 210 | 224 | 239 | 254 | 269 | 284 | 298 |
| Hard coal | | | | | | | | | | |
| IC (€/kW) | 3238 | 3303 | 3369 | 3435 | 3500 | 3566 | 3632 | 3697 | 3763 | 3828 |
| FOM (€/kW/a) | 44 | 45 | 47 | 49 | 50 | 52 | 54 | 55 | 57 | 59 |
| AIC (€/kW/a) | 285 | 292 | 298 | 305 | 311 | 318 | 324 | 331 | 338 | 344 |
| CCGT | | | | | | | | | | |
| IC (€/kW) | 889 | 898 | 907 | 916 | 925 | 934 | 943 | 953 | 962 | 971 |
| FOM (€/kW/a) | 13 | 14 | 15 | 16 | 16 | 17 | 18 | 19 | 20 | 21 |
| AIC (€/kW/a) | 84 | 85 | 87 | 89 | 90 | 92 | 94 | 95 | 97 | 99 |
| OCGT | | | | | | | | | | |
| IC (€/kW) | 634 | 653 | 672 | 690 | 709 | 727 | 746 | 765 | 783 | 802 |
| FOM (€/kW/a) | 11 | 11 | 11 | 12 | 12 | 12 | 12 | 12 | 12 | 12 |
| AIC (€/kW/a) | 62 | 63 | 65 | 67 | 68 | 70 | 71 | 73 | 75 | 76 |
| Solar PV | | | | | | | | | | |
| IC (€/kW) | 1421 | 1345 | 1269 | 1193 | 1117 | 1041 | 965 | 890 | 814 | 738 |

| FOM (€/kW/a) | 22 | 21 | 20 | 19 | 18 | 17 | 16 | 15 | 14 | 13 |
|---|---|---|---|---|---|---|---|---|---|---|
| AIC (€/kW/a) | 142 | 135 | 127 | 120 | 112 | 105 | 97 | 90 | 82 | 75 |
| Wind Onshore | | | | | | | | | | |
| IC (€/kW) | 1336 | 1352 | 1368 | 1384 | 1400 | 1417 | 1433 | 1449 | 1465 | 1481 |
| FOM (€/kW/a) | 41 | 39 | 37 | 36 | 34 | 32 | 30 | 28 | 26 | 24 |
| AIC (€/kW/a) | 165 | 165 | 165 | 164 | 164 | 164 | 163 | 163 | 162 | 162 |
| Wind Offshore | | | | | | | | | | |
| IC (€/kW) | 3564 | 3514 | 3465 | 3415 | 3365 | 3316 | 3266 | 3216 | 3166 | 3117 |
| FOM (€/kW/a) | 103 | 100 | 96 | 93 | 89 | 85 | 82 | 78 | 75 | 71 |
| AIC (€/kW/a) | 405 | 397 | 390 | 382 | 374 | 366 | 359 | 351 | 343 | 335 |
| Bioenergy | | | | | | | | | | |
| IC (€/kW) | 3212 | 3366 | 3521 | 3676 | 3831 | 3986 | 4140 | 4295 | 4450 | 4605 |
| FOM (€/kW/a) | 97 | 101 | 106 | 110 | 114 | 119 | 123 | 128 | 132 | 136 |
| AIC (€/kW/a) | 353 | 370 | 386 | 403 | 420 | 437 | 453 | 470 | 487 | 504 |
| Geothermal | | | | | | | | | | |
| IC (€/kW) | 7751 | 7993 | 8234 | 8476 | 8717 | 8958 | 9200 | 9441 | 9683 | 9924 |
| FOM (€/kW/a) | 339 | 341 | 289 | 265 | 240 | 215 | 190 | 165 | 141 | 116 |
| AIC (€/kW/a) | 917 | 910 | 903 | 896 | 889 | 883 | 876 | 869 | 862 | 855 |
| Battery Storage | | | | | | | | | | |
| €/kW | 1891 | 1857 | 1823 | 1789 | 1755 | 1721 | 1687 | 1653 | 1619 | 1585 |
| €/MWh | 803 | 736 | 668 | 601 | 534 | 467 | 400 | 332 | 265 | 198 |
| AIC (€/kW/a) | 246 | 242 | 238 | 233 | 229 | 224 | 220 | 215 | 211 | 207 |

Table 5 presents the assumed investment costs, fixed operation and maintenance costs, and annualised investment costs for the cost-optimal expansion in the future analysis.

**Table 5. Future cost data**

| | **2030** | **2035** | **2040** | **2045** | **2050** |
|---|---|---|---|---|---|
| Solar PV | | | | | |
| IC (€/kW) | 530 | 389 | 286 | 210 | 154 |
| FOM (€/kW/a) | 9 | 8 | 6 | 5 | 4 |

| AIC (€/kW/a) | 54 | 41 | 31 | 23 | 18 |
|---|---|---|---|---|---|
| Wind Onshore | | | | | |
| IC (€/kW) | 1278 | 1217 | 1160 | 1105 | 1053 |
| FOM (€/kW/a) | 20 | 16 | 13 | 10 | 8 |
| AIC (€/kW/a) | 129 | 119 | 111 | 104 | 97 |
| Wind Offshore | | | | | |
| IC (€/kW) | 2755 | 2566 | 2390 | 2226 | 2074 |
| FOM (€/kW/a) | 58 | 48 | 39 | 32 | 26 |
| AIC (€/kW/a) | 292 | 265 | 241 | 221 | 202 |
| Battery Storage | | | | | |
| €/kW | 1233 | 1043 | 881 | 745 | 630 |
| €/MWh | 186 | 164 | 99 | 60 | 36 |
| AIC (€/kW/a) | 161 | 136 | 115 | 97 | 82 |

# Appendix D. Fuel and $CO_2$ Price Data

Table 6 shows fuel and $CO_2$ price data for the historical period with source references. Our analysis requires fuel prices for uranium, lignite, hard coal, and natural gas. The historical fuel prices for all technologies in both systems are based on the literature [3,71,88,87,89,90], with significant differences occurring from Germany to Texas during the gas crisis in 2022. Fuel prices in the future time period are also extrapolated linearly to 2050. The same approach is also used for the $CO_2$ price, which is listed for only Germany, as no emission trading system has been established in Texas. For the historical time period, the annual average price of the EU ETS applies to Germany, based on [82]. For the future time periods, the ambitious $CO_2$-price scenario from [31] is adopted. In contrast, Texas has no existing $CO_2$ pricing system. Therefore, Texas remains for both periods without a $CO_2$ pricing system.

**Table 6. Fuel and $CO_2$ price data for the historical period with source references**

| | 2015 | 2016 | 2017 | 2018 | 2019 | 2020 | 2021 | 2022 | 2023 | 2024 | REF |
|---|---|---|---|---|---|---|---|---|---|---|---|
| **Fuel price (€/$MWh_{th}$)** | | | | | | | | | | | |
| DE & TX | | | | | | | | | | | |
| Uranium | 1.67 | 1.76 | 1.91 | 2.02 | 2.07 | 2.11 | 2.22 | 2.36 | 2.5 | 2.61 | [91] |
| Lignite | 1.60 | 1.59 | 1.55 | 1.57 | 1.65 | 1.67 | 1.71 | 1.80 | 2.24 | 2.68 | [89,90] |
| DE | | | | | | | | | | | |
| Hard coal | 11.16 | 10.33 | 16.08 | 16.19 | 14.74 | 11.21 | 15.45 | 39.10 | 26.07 | 21.90 | [89,90] |
| Natural gas | 28.03 | 16.59 | 21.07 | 27.09 | 15.77 | 10.94 | 44.33 | 127.75 | 41.41 | 35.99 | [88,89] |
| TX | | | | | | | | | | | |
| Hard coal | 6.97 | 6.56 | 6.84 | 7.07 | 7.16 | 7.28 | 7.35 | 7.41 | 7.48 | 7.53 | [91] |
| Natural gas | 10.72 | 9.33 | 10.60 | 10.45 | 8.82 | 7.06 | 14.49 | 20.07 | 10.89 | 17.16 | [3] |
| **$CO_2$ price (€/t$CO_2$)** | | | | | | | | | | | |
| EU ETS | 7.5 | 8 | 5.8 | 15 | 24.9 | 24.8 | 52.9 | 79.7 | 83.3 | 67.5 | [82] |

Table 7 reflects the fuel and carbon price assumptions for the future time period.

**Table 7. Fuel and $CO_2$ price assumptions for the future time period**

| | 2030 | 2035 | 2040 | 2045 | 2050 |
|---|---|---|---|---|---|
| ***Fuel price (€/$MWh_{th}$)** | | | | | |
| DE & TX | | | | | |
| Uranium | 3.23 | 3.74 | 4.25 | 4.75 | 5.26 |
| Lignite | 3.10 | 6.68 | 4.27 | 4.85 | 5.44 |
| DE | | | | | |
| Hard coal | 26.30 | 31.40 | 36.50 | 41.60 | 46.71 |
| Natural gas | 32.63 | 37.15 | 41.68 | 46.20 | 50.72 |
| TX | | | | | |
| Hard coal | 8.16 | 8.61 | 9.10 | 9.52 | 9.97 |
| Natural gas | 17.55 | 20.59 | 23.63 | 26.67 | 29.72 |
| ****$CO_2$ price (€/t$CO_2$)** | | | | | |
| EU ETS | 130 | 170 | 210 | 270 | 350 |

*Linear extrapolation without consideration of crisis years 2021, 2022 and 2023

**EU ETS future price assumptions from [31]

# Appendix E. Reference Weather Year

The fluctuations in the annual average capacity factors from 1980 to 2024 are shown in Figure 7. The division into different regions according to section 3.2.2 applies here. Texas' more favourable environmental conditions compared to Germany's are apparent. Solar is almost twice as high, at up to 26%, and onshore wind is around 10% higher than in Germany. However, regarding offshore wind, Germany's high potential is evident due to the high wind speeds at sea. In Texas, these are significantly lower compared to Germany and to the inland wind potential. The wind maps for both systems can be found in Appendix F.

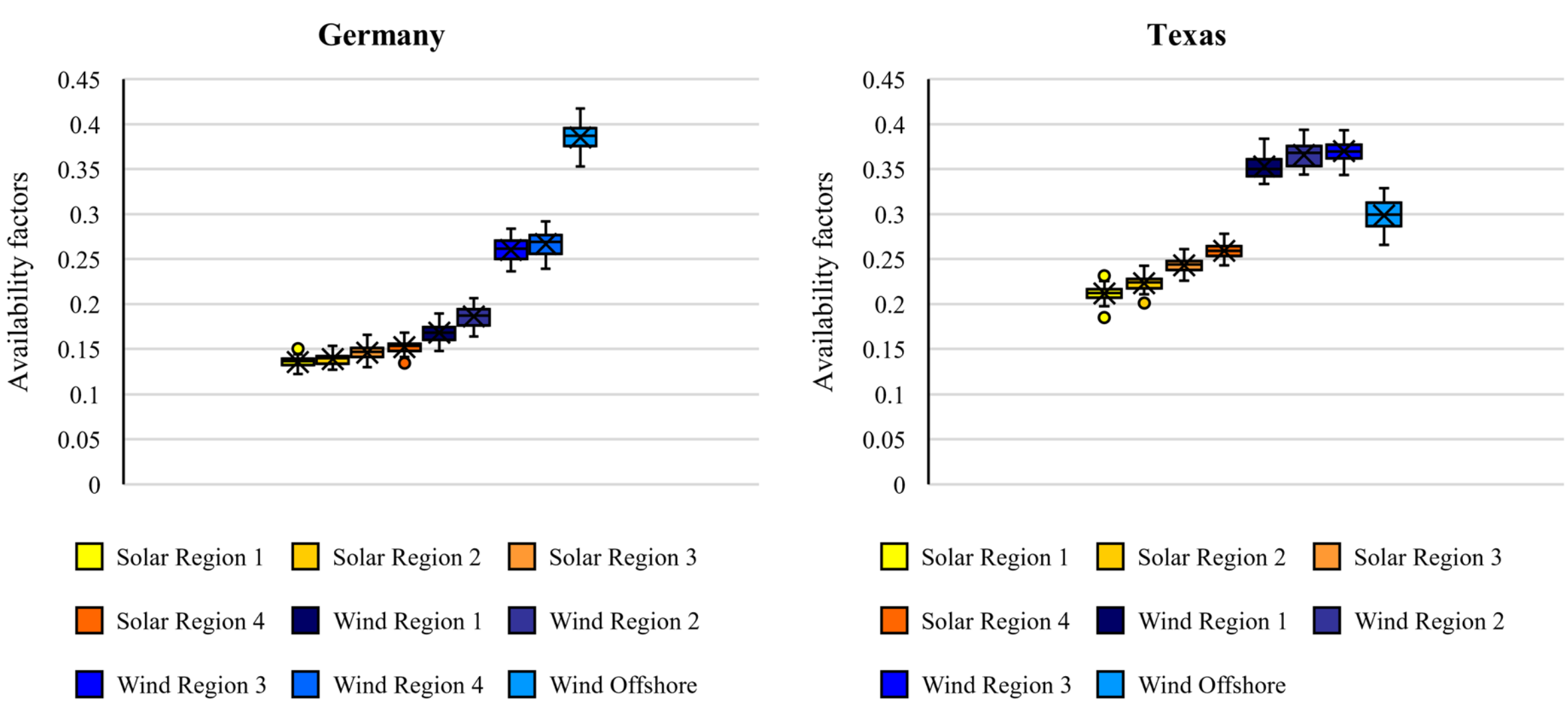

**Figure 7. Annual average capacity factors from 1980 to 2024 for Germany and Texas by solar and wind regions**

In order to minimise the influence of weather conditions, a reference weather year is developed for our future perspective in section 4.2, based on the median years by region in Table 8. The hourly capacity factors for the respective year in the relevant region represent the model input. The sensitivity analysis of the reference weather year can be found in Appendix G.

**Table 8. Median years of average capacity factors by solar and wind regions for Germany and Texas according to Figure 7**

| Median year for solar regions (1980-2024) | | | | Median year for wind regions (1980-2024) | | | | |
|---|---|---|---|---|---|---|---|---|
| Region 1 | Region 2 | Region 3 | Region 4 | Region 1 | Region 2 | Region 3 | Region 4 | Offshore |

<table>
<tr><th colspan="9">Germany</th></tr>
<tr><td>1989</td><td>1994</td><td>2002</td><td>1989</td><td>1981</td><td>1999</td><td>2020</td><td>2022</td><td>2023</td></tr>
<tr><th colspan="9">Texas</th></tr>
<tr><td>2001</td><td>2017</td><td>2005</td><td>2002</td><td>2022</td><td>2001</td><td>1998</td><td>-</td><td>1993</td></tr>
</table>

# Appendix F. Comparison of Wind Speeds for Germany and Texas

With regard to our historical perspective in section 4.1 and our future perspective in section 4.2, on the one hand, it is evident that Germany is experiencing significant growth in offshore wind generation. Texas, on the other hand, is only expected to expand its onshore wind capacity in the future, while offshore wind remains absent from the market until 2050. In order to better understand the potentials of wind and thus also the results, we refer to a wind speed map from [92] for Germany and Texas. Wind speeds are recorded for offshore wind in Germany (North Sea) and on- and offshore (Gulf of America) wind in Texas in Appendix E. A direct comparison reveals the high potential of offshore wind in Germany, with a mean wind speed of 10 m/s. The map also visualises the higher potential compared to onshore wind. In contrast, it is the opposite in Texas. Here, we have a higher mean wind speed onshore at 9 m/s than offshore at 7.5 m/s. Comparatively less favourable conditions and higher investment costs of offshore wind result in a lack of competitiveness and thus a lack of expansion in our perspectives for Texas.

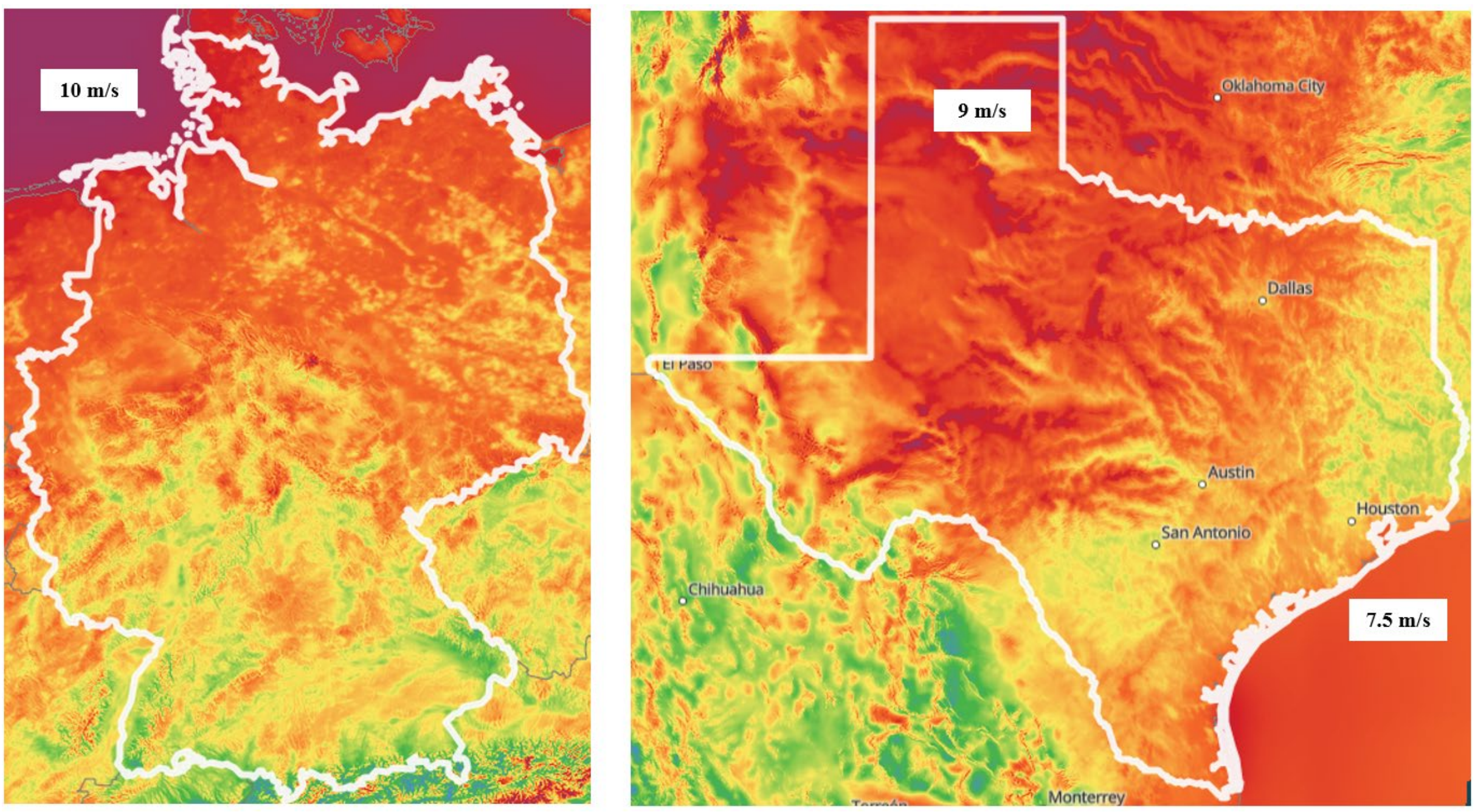

**Figure 8. Wind speed map for Germany (offshore) and Texas (on- and offshore) according to the Global Wind Atlas**

## Appendix G. Sensitivity Analysis

In addition to analysing the annual historical expansion and the individual effects from 2015 to 2024, we also perform a sensitivity analysis. Here we set the parameter values to the average of the historical period (except weather) and keep them at a constant level. This enables us to consider cost-optimal electricity systems independently of the largest influencing factors. Our results refer to the deviation from the historical analysis in section 4.1. Below, a positive value represents an increase and a negative value a decrease in our sensitivity analysis compared to the historical results. This is intended to provide the reader with further insights into the model results for deviating data assumptions.

For the sensitivity analysis of the weather, we used the reference years for the respective solar and wind regions from Table 8 in our historical observation period. The difference in the annual average capacity factors is shown on the right-hand side in Figure 9, and the deviations from the historical perspective are illustrated on the left hand-side for Germany and Texas. The model results show that the weather only has an influence when solar PV and wind turbines are also part of the system. This means that the deviations after 2021 remain below 20% for Germany and below 5% for Texas. Since Texas has only installed solar PV capacity in our historical perspective, the deviations of the total generation share are minimal. In Germany, the year 2023 shows the largest deviation, with onshore wind being 17% lower in contrast to the historical perspective due to the lower-capacity factors, which is almost entirely offset by offshore wind with 14%. Furthermore, 2024 shows that lower-capacity factors of wind result in a higher share of CGCT (7%) and solar PV (3%). In conclusion, we can see an increasing shift toward generation from solar PV and wind turbines.

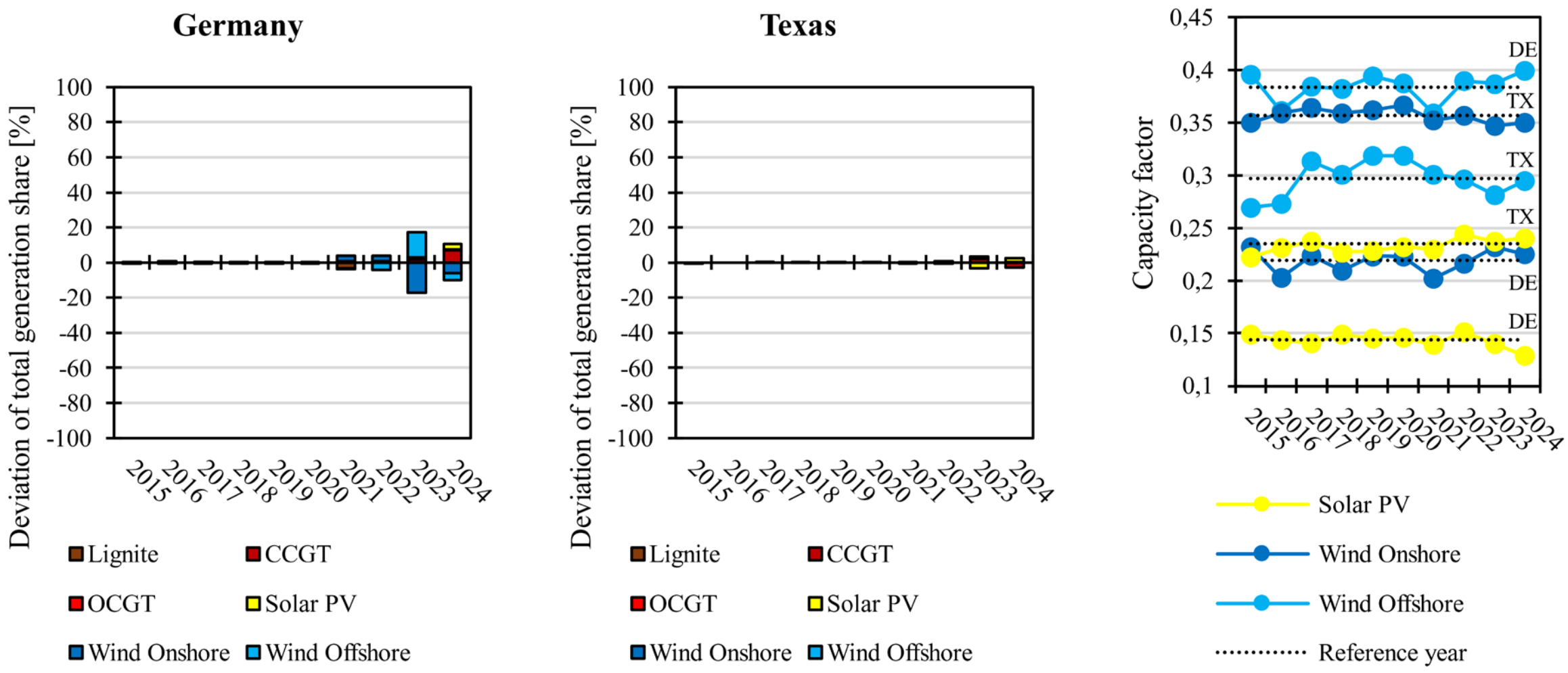

**Figure 9. Sensitivity analysis for reference weather year in relation to the historical perspective**

The basis of the sensitivity analysis of the $CO_2$ price reflects the fixed price of 36.9 €/ton for all years under consideration in Germany. All other parameters remain at their annual values. As previously mentioned, Texas does not have an emission trading system, hence the carbon price is 0 €/ton and an analysis would not show any deviations from Figure 2. For this reason, the sensitivity analysis is only carried out for Germany. The annual deviation of total generation share and the average carbon price are depicted in Figure 10.

In the initial phase from 2015 to 2020, the higher $CO_2$ price reduces the competitiveness of lignite and gas-fired plants in comparison to solar PV and wind. This highlights two facts. On the one hand we can recognise the total shift from lignite to gas-fired plants in 2016 and 2017. Due to the higher carbon content of lignite, gas-fired power plants are now the main producer in the cost-optimal system for these years. The gas shift does not occur in 2015 nor in 2018, as the fuel prices for natural gas were significantly higher, compared to the other years, at 27 €/MWh to 28 €/MWh. This shows a strong dependency on the local gas price. On the other hand, despite the higher market entry potential of solar PV and wind turbines, the electricity system still operates almost entirely with conventional sources. In 2017, onshore wind accounts for approximately 1.4%, which decreases to 0.7% in 2018, while solar PV accounts for 0.06%. This shows the competition of gas-fired power plants on the market, which can replace renewables in expansion with a low carbon and natural gas price.

In the following phase from 2021 to 2024, a significantly higher proportion of lignite is expected. First, our average carbon price is below the real one. Second, the sharp rise in gas prices caused by the conflict with Russia is having an impact. That results in a higher competitiveness of lignite power plants and an overall increase in the conventional generation share, as compared to the historical perspective. However, solar PV as well as wind turbines remain a part of the cost-optimal system. An exception to this is offshore wind, which is fully compensated. In 2021, the remaining RES share is 8%, and in 2022 it increased to 13%. Due to the high gas price peak of 128 €/MWh in 2022, the demand is still mostly covered by lignite power plants, but a higher RES share becomes apparent. With a look at our peak renewable expansion year 2023, despite a decrease of -10% in solar PV and -22% in onshore wind generation, the total share increases to a level of 16%. For 2024, the remaining RES share in sum is around 18%. Despite the low $CO_2$ price and natural gas price fluctuations, a clear upward trend in generation from solar PV and wind can be recognised.

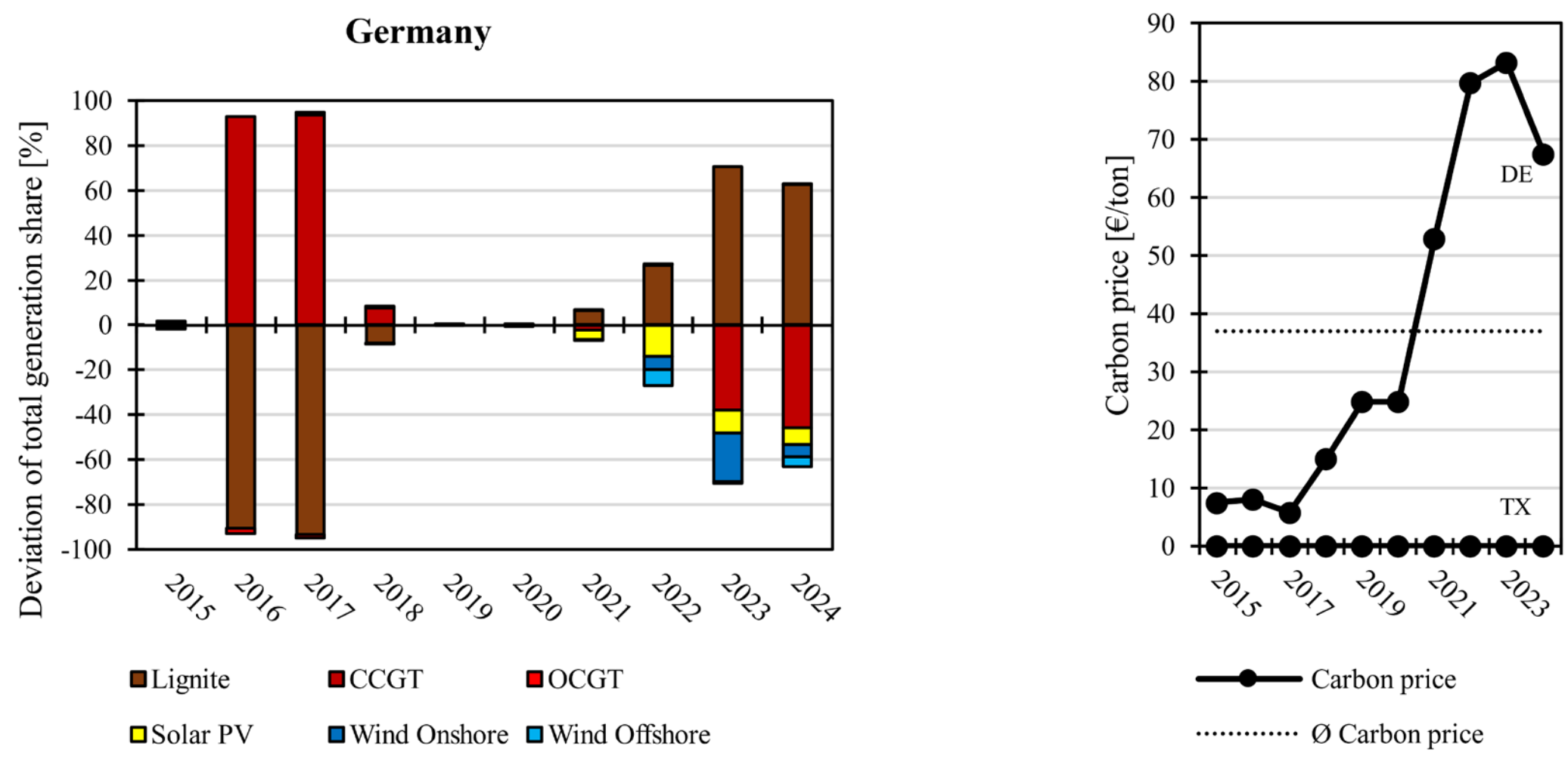

**Figure 10. Sensitivity analysis for average carbon price in relation to the historical perspective**

The results of the sensitivity analysis for fuel prices are presented in Figure 11. We have set the costs for uranium, lignite, hard coal, and natural gas at the average value from 2015 to 2024. The average fuel prices for uranium are 5.88 €/MWh, and for lignite, 1.80 €/MWh. Both systems have the same price due to the limited data availability and the local mining of lignite. There is a difference in the average price of hard coal between Germany and Texas, at 18.22 €/MWh and 7.31 €/MWh, respectively. In addition, separate natural gas prices are also considered. The average price in Germany is 36.90 €/MWh, and 12 €/MWh in Texas. Furthermore, natural gas in particular shows strong fluctuations with a peak in 2022 for both systems, as can be seen in the right-hand part of the figure. In this regard, other fuel prices remain relatively constant, which is why only the natural gas price is reflected here. The left-hand side of the figure clearly shows in both systems that the higher average gas price up to 2020 leads to a switch to lignite-based generation. Onshore wind enters the market in Germany with a 3.6% share, and solar PV in Texas with a 0.58% share, as early as 2020, instead of in 2021. Due to the low natural gas price in 2020 for both systems, the higher average value in this analysis ensures that they are becoming more competitive. After 2020, the average costs are below the real costs, which reduces the variable costs of gas-fired power plants. Despite this, solar PV and wind turbines remain in the cost-optimal system, decreasing only slightly. In Germany, onshore wind turbines are replaced in 2023 with -9.17%, and in Texas, solar PV with -2.52% in 2024 from CCGT. In summary, the increasing competitiveness of solar PV and wind turbines is evident even at a low natural gas price.

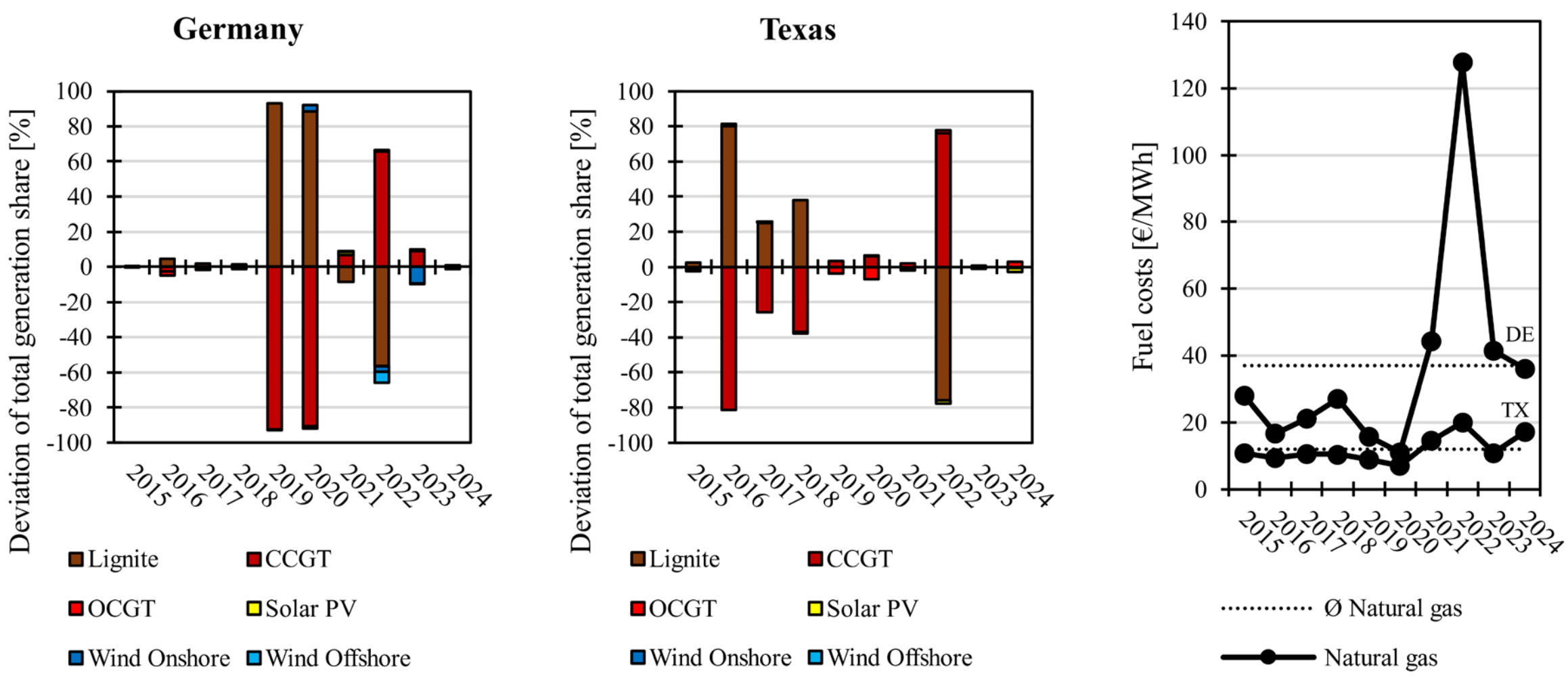

**Figure 11. Sensitivity analysis for average fuel prices in relation to the historical perspective**

As we have discussed in section 4.3, the development of investment costs is an important influencing factor for Germany and Texas. We set the investment costs to an average value using the capital and fixed operation and maintenance costs. This applies to both conventional and renewable technologies. The yearly deviations in the historical perspective are shown in Figure 12 for Germany and Texas.

First, we focus again on the initial phase up to 2020. In Germany, only minor deviations are recognisable, except for 2016, when 20% of lignite generation is replaced with CCGT. This is attributable to the decreasing gas price, in contrast to the lignite-based years in 2015, 2017 and 2018. However, this is also confirmed by the higher increase of capital costs in lignite power plants at 2520 €/kW, and gas-fired plants of CCGT at 930 €/kW. We find similar causation in the Texan system in 2015 and 2017, because lignite generation is completely replaced by CCGT at almost constant natural gas prices. Despite the lower investment costs of solar PV and wind systems in this phase, there is no earlier expansion of these technologies in either system.

The following phase shows strong deviations from the historical perspective, but also between the two systems. In Germany, deviations occur especially in the energy crisis years 2022 and 2023. It is particularly worth emphasising here that conventional generation is covered mostly by nuclear power plants, regardless of the fuel. In 2022 this leads to a decreasing share of solar PV at -11.5%, onshore wind at -4%, and offshore wind is completely replaced at -6.8% in the previously lignite-based system. This brings the renewable share to 17%. The switch to a nuclear-based system is also recognisable in 2023. However, here the previously installed CCGTs are replaced. The total share of renewable generation is 19%. The lower investment costs, the sharp rise in the carbon price, and the sharp rise in the natural gas price ensure the increasing competitiveness of nuclear power plants and therefore are also part of the cost-optimal system. Nevertheless, it is also clear that the expansion of nuclear power plants reduces the share of renewables. This leads to the assumption that without the German government's decision to phase out nuclear power in 2011, the renewable expansion potential would have been lower afterwards. In addition, two more facts are confirmed by the 2024 results for Germany. First, the share of solar PV and wind energy decreases from 35% in the historical perspective to 17.5% in our investment cost sensitivity analysis, whereby this remaining share basically corresponds to onshore wind. Second, the shares from solar PV and offshore wind are covered by CCGT, as in the previous $CO_2$ analysis. Both results lead to the assumption that onshore wind in particular plays an important role in the interaction with gas-fired plants in the German system. The following findings can be determined for Texas in the second phase: the lower investment costs of lignite power plants and rising natural gas price results in an additional lignite-based system in 2021 and 2024. With regard to renewables, the share of solar PV decreases in all years, due to their higher investment costs in this phase. However, a small share still remains in the lignite-based years, such as 2021 and 2022 with 1.8% each, and 2024 with 3.9%. In 2023, with low natural gas prices immediately after the crisis, solar PV disappears from the energy system completely. This again reflects the competitiveness of gas-fired power plants, here with solar PV in Texas.

Overall, the sensitivity analysis of investment costs also confirms that the phase after 2020 is primarily important for the renewable expansion. Although there is no earlier market entry in the first phase, in the second phase neither solar PV nor onshore wind disappear from the cost-optimal system in either system. Despite the lower potential of renewables after 2020 due to their high investment costs and the falling costs for conventionals in this analysis, solar PV in Texas and additional onshore wind in Germany remain as competitive technologies in the energy system.

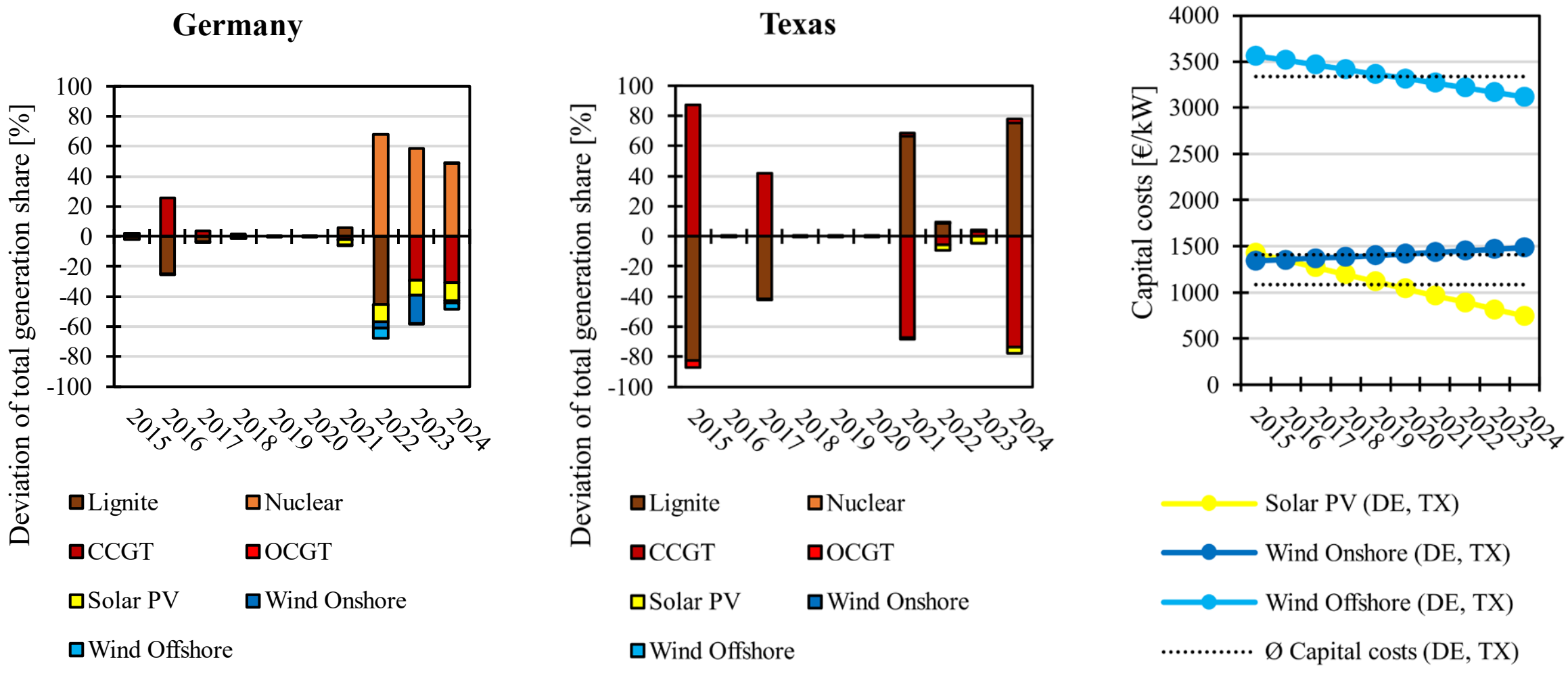

**Figure 12. Sensitivity analysis for average investment costs in relation to the historical perspective**

In the following, we set the hourly electricity demand to an average hourly value. Figure 13 displays the results, and the annual demand for Germany and Texas. In Germany, the deviations remain below 10%, although these occur after 2021. Due to the sharp decline in demand after 2021, the average is now significantly above the historical value. In addition, averaging smooths out peak loads, which favours base load power plants like lignite and CCGT. As a result, the share of conventional generation in Germany is also higher. In particular, generation from wind turbines is replaced by lignite-fired power plants in 2021 (5.14%), and by CCGT in 2023 (5.12%) and 2024 (3.73%). Solar PV shows an increase of 1% in each of these years. In contrast to Germany, Texas has increased its electricity demand from 350 TWh in 2015 to approximately 460 TWh in 2024. The highest deviation of our demand analysis occurs in 2017, when 10% of CCGT generation is replaced by lignite, which also corresponds to the lower peaks in demand. With regard to 2024, Texas reports a 2.1% higher share of solar PV generation, which consequently results in a decrease of gas generation. Both systems allocate a fixed share of renewable generation to both higher demand (DE) and lower demand (TX). Solar PV in particular is emphasising its competitiveness through its growth in share.

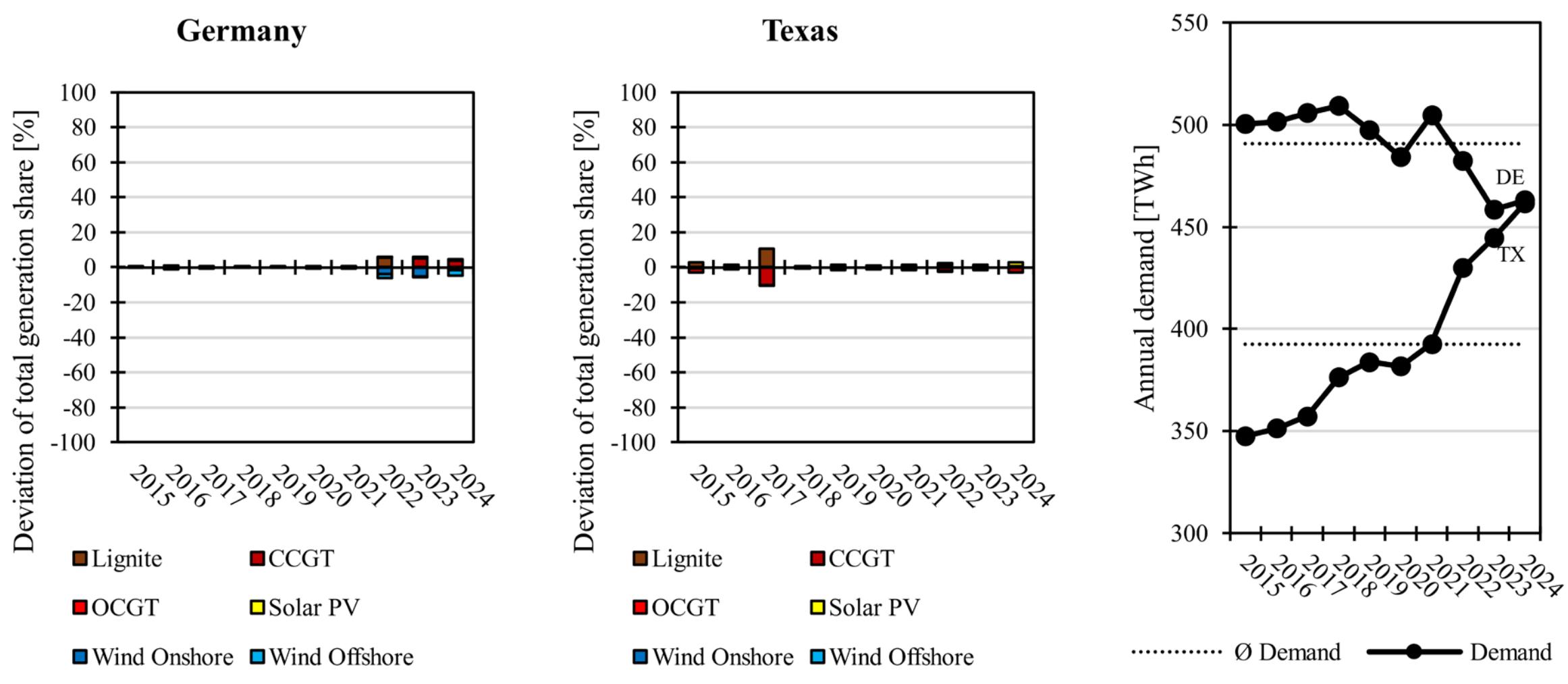

**Figure 13. Sensitivity analysis for average electricity demand in relation to the historical perspective**

The last parameter to be considered is the cross-border flow, which is implemented via the hourly net export. Figure 14 illustrates the deviations from the historical perspective on the left as well as the annual cumulative exports on the right for Germany and Texas. The deviations for Texas are less than 1% across all years, as the ERCOT network operates in reality almost as an island. In contrast, Germany has a strong trade relation with its neighbouring countries, but has developed from an exporter into an importer by 2022. By summing up the exogenous demand and cross-border flow in the model, results similar to those shown in the sensitivity analysis for demand in Figure 13 are obtained. In this case, wind generation in Germany is replaced by lignite (2022) and CCGT (2023, 2024). In 2023 alone, the onshore wind share decreases by -14.14%. This reduction can be attributed to the higher average trading volume in the historical perspective, as well as less volatile trade in the system. Despite these reductions, solar PV and onshore wind remain as a part of the cost-optimal system.

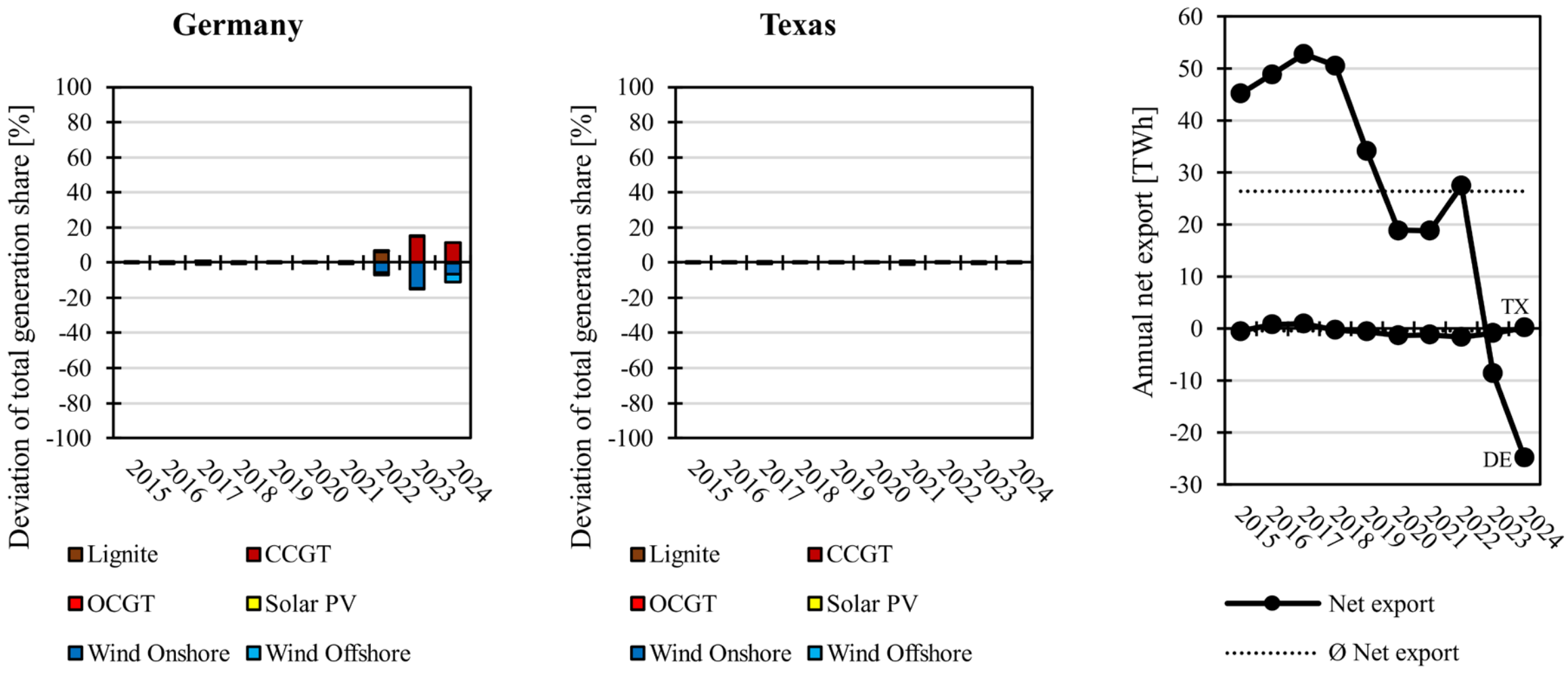

**Figure 14. Sensitivity analysis for average cross-border flow in relation to the historical perspective**

## Appendix H. Limited Scenario: Annual Cost-Optimal Expansion

We quantify the impact of realistic expansion limitations on the results as compared to the historical perspective in Figure 15. The high generation share of lignite power plants and the resultingly high installed capacities raise questions about the feasibility of this scenario in reality. With a view to a cost-optimal system without restrictions, the historical results are certainly possible. However, in this section we implement upper capacity limits for lignite, nuclear, and hard coal power plants for each system. The limits of generation capacities refer to the maximum values from the past from [70,73] and are visualised on the right-hand side of the figure. Other technologies remain open for expansion. With regard to Germany, we achieve a higher share of approximately 65% of CCGT by 2018 due to the restrictions of lignite capacity. In Texas, the same effect becomes visible in 2015, 2017, and 2022, while the other years remain unchanged. Further deviations are evident in Germany during the crisis years of 2021 and 2022. With the rise in $CO_2$ price and the restriction of lignite capacity, the system is increasingly building CCGT, as in previous years. In addition, gas prices are rising, which further decreases the cost competitiveness of gas-fired plants. This is also reflected in the 5% increase in solar PV and 3.5% in onshore wind turbines. With gas prices peaking in 2022 and $CO_2$ prices rising even higher, Germany ends up with a nuclear-based system. This also results in a 3.5% and 3.8% increase in onshore and offshore wind respectively, while solar PV decreases by 3.9%. Texas also sees a record high natural gas price in 2022, causing solar PV to achieve a 3% higher share. This highlights two important facts. First, high natural gas prices combined with high $CO_2$ (for Germany) could lead to nuclear power plants becoming competitive again. Second, the generation mix clearly shows that nuclear power plants tend to operate with wind turbines and exclude solar PV. Texas underlines that solar PV performs better with gas-fired power plants. Overall, the phase-out of nuclear power in Germany has promoted the expansion of RES, but in the future, rising prices for $CO_2$ and natural gas could lead to a renewed competitiveness that is purely cost-based. The same effects must be taken into account for Texas.

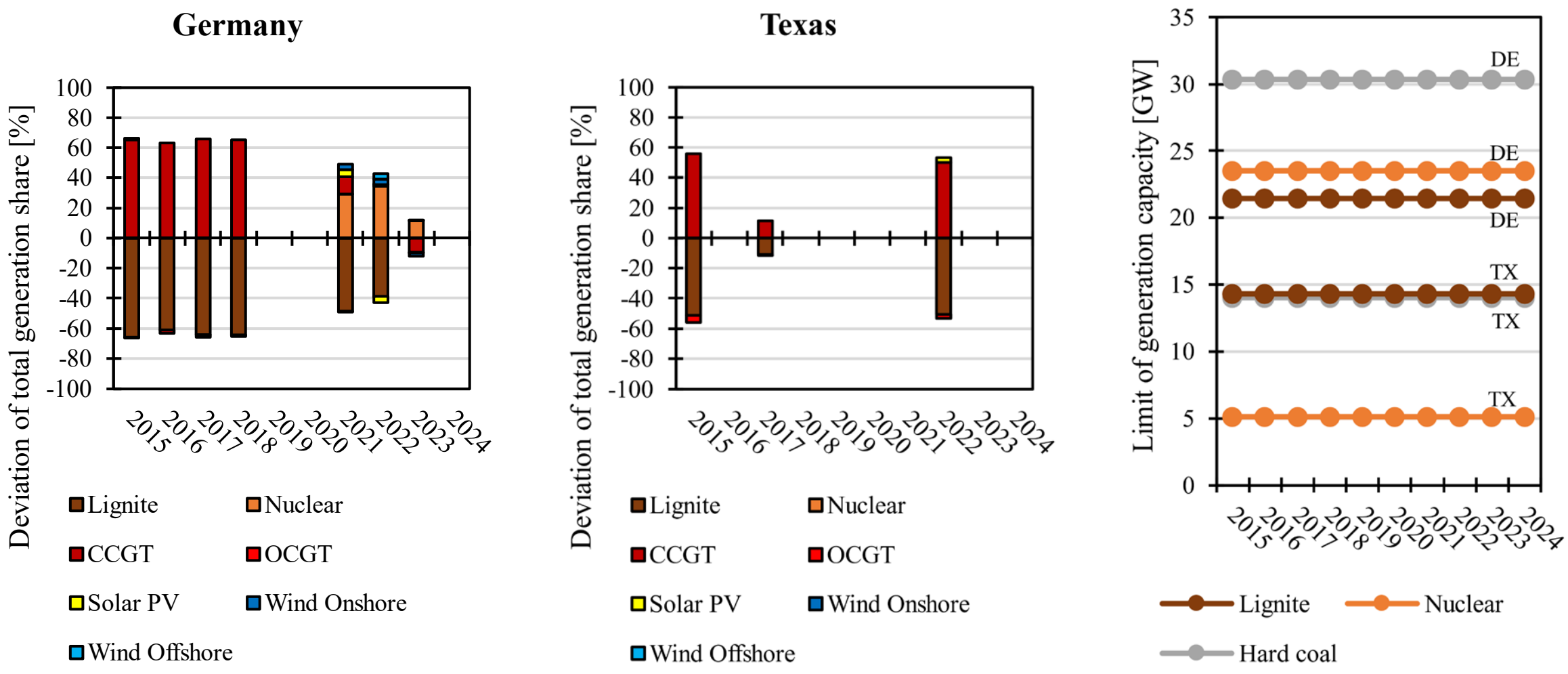

**Figure 15. Limited scenario with upper capacity limitations in relation to the historical perspective**